\documentclass[a4paper,10pt]{article}
\usepackage{jheppub} % for details on the use of the package, please see the JINST-author-manual
\usepackage{orcidlink}
\newcommand{\nn}{\nonumber}
\usepackage{amsmath,amssymb}
\usepackage[toc,page]{appendix}
\usepackage{comment}
\usepackage{hyperref}
\usepackage{mathrsfs}
\usepackage{mathtools}
\usepackage{empheq}
\arxivnumber{2407.08379} % if you have one
\usepackage{academicons}
\title{\boldmath Light transformation: A Celestial and Carrollian perspective
}

\author[1]{Sourish Banerjee \orcidlink{0000-0003-2618-1025}}
\author[1]{Rudranil Basu \orcidlink{0000-0003-0655-0890}}
\author[2]{Sayali Atul Bhatkar}
\affiliation[1]{Department of Physics, Birla Institute of Technology and Science Pilani, \\
Zuarinagar, Goa 403726, India}
\affiliation[2]{Intelligent Systems group Bernoulli Institute University of Groningen, \\ 
P.O. Box 407 9700 AK Groningen, The Netherlands}
\emailAdd{p20210001@goa.bits-pilani.ac.in}
\emailAdd{rudranilb@goa.bits-pilani.ac.in}
\emailAdd{s.bhatkar@rug.nl}
\abstract{In this paper, we first study the consequence of spacetime translations and Lorentz transformations on Celestial CFT OPEs. Working with the light transforms of the operators belonging to the modified Mellin basis,  we found that the leading order singularity in the OPE of such operators could be fixed purely using Poincar\'e symmetries owing to the non-trivial action of the translations on these operators. The OPE coefficient is then fixed using the soft limit of the correlation functions. We check that this singular structure obtained from symmetries is consistent with the OPE limit of three-point functions. This approach could potentially be useful for studying Celestial CFT without adverting to bulk physics. As another goal, we explore the significance of light transformation in Carrollian CFTs. In the special cases we considered, we show that light transformation equips us with a map between two branches of Carroll CFT in $d=3$ dimension at the level of correlation functions in the near coincident limit. }

\begin{document}
\maketitle
\flushbottom

\section{Introduction}
\label{Introduction}
The computation of tree-level gluon and graviton scattering amplitudes in flat space has been mapped to calculating correlation functions of conformal field theories on the celestial sphere in the past decade\cite{Pasterski:2017ylz, Pasterski:2016qvg, Schreiber:2017jsr, Cheung:2016iub, Banerjee:2017jeg}. Unlike usual 2D conformal field theories (CFT), celestial conformal field theories (CCFT) are constrained by further symmetries: translations along the null-time direction, orthogonal to the celestial sphere. This enlarges the global symmetry group from Lorentz (conformal group of the 2-sphere) to Poincar\'{e} \cite{Banerjee:2018gce, Banerjee:2020kaa, Stieberger:2018onx, Law:2019glh}. As a result, the correlation functions of local operators become distributional in nature. This is consistent with bulk computations, as the Dirac delta function behaviour reflects momentum conservation of bulk amplitudes. However, owing to this behaviour of correlation functions, local data like operator product expansion (OPE) coefficients are not straightforward to obtain in these theories from the collinear limit of three-point functions as one would expect from the perspective of a 2D CFT \cite{DiFrancesco:1997nk}. Although OPEs in CCFT have been studied, the analysis starts with the four-point function \cite{Pate:2019lpp, Banerjee:2020vnt, Banerjee:2020kaa, Ren:2023trv, De:2022gjn, Hu:2022syq, Adamo:2022wjo, Bhardwaj:2022anh, Tao:2024rlo} due to the presence of Dirac delta singularities in three-point functions. 

This apparent problem stems from the translation symmetry of the 4D theory. This symmetry introduces an additional structure to CCFTs that is different from ordinary CFTs. In this paper, we start with the question of whether it is possible to exploit the translation symmetry of the theory to further constrain the putative boundary conformal theory. This is desirable as this would be an independent boundary calculation without using any bulk information. As a definite step in this direction, we show that the leading singularity in OPE of gluons is completely fixed up to a numerical factor. It should be noted that conformal symmetry dictates that OPEs take a power law behaviour. We show that the exact power of this singularity in the OPE limit of two gluons can be fixed using translation symmetry. In order to demonstrate this, we need to change the basis as obtaining the OPE limit of three-point functions in the usual celestial basis is difficult.

The reason behind the difficulty in getting OPE from % As we mentioned, it is difficult to obtain an OPE from 
the collinear limit of the Celestial CFT three-point function is two-fold: 1) the three-point function is ultra-local and one gets contact terms in the OPE of such operators and 2) the three-point function in Celestial CFT is heavily constrained on the Celestial sphere due to bulk momentum conservation, giving rise to the following subtlety. If we focus on the $12 \leftrightharpoons3$ channel of the three-point function in Mellin basis $ \langle\phi_{(h_{1},\Bar{h}_{1})}(z_{1},\bar{z}_{1})\phi_{(h_{2},\Bar{h}_{2})}(z_{2},\bar{z}_{2})\phi_{(h_{3},\Bar{h}_{3})}(z_{3},\bar{z}_{3})\rangle $, then we see that it is non-zero only when the points at which the fields are inserted are ordered as $z_{1}<z_{3}<z_{2}$ or $z_{1}>z_{3}>z_{2}$. So, even though one can take the collinear limit where $z_{1}\rightarrow z_{2}$ one can not call this the OPE since this limit, due to non-trivial support of the three-point function, effectively brings all the three points close to one another. One would then be compelled to study the $z_{2}\rightarrow z_{3}$ (or $z_{1}\rightarrow z_{3}$) limit of the three-point function, which averts this issue, only to end up with delta function singularities\cite{Banerjee:2022hgc}. It is imperative that if one chooses a basis of operators that are smeared on the celestial sphere, one may get rid of the contact terms in leading order, and an unambiguous OPE can be obtained. Hence we are motivated to study the light-transformed operators \cite{Sharma:2021gcz, Crawley:2021ivb, Pasterski:2017kqt}. The integral transform inherently involved in defining these non-local operators rips off the distributional or ultra-local nature of the algebra of these operators \cite{De:2022gjn, Banerjee:2022hgc, Hu:2022syq, Furugori:2023hgv, Chang:2022jut, Crawley:2021ivb} and consequently of the correlation functions as well. OPEs of such operators have been found in \cite{Sharma:2021gcz, De:2022gjn, Hu:2022syq}, and it was shown in \cite{Banerjee:2022hgc} that the OPEs obtained from the three-point function had non-trivial channel dependence at the level of the OPE coefficients which needs to be studied further.
\begin{comment}   
For example, the light ray transform (considering Lorentzian signature in the CCFT) of a scalar Mellin primary $O_{\Delta}$ is
\begin{eqnarray}
    L^{+}[{O}_{\Delta}](z,\bar{z})=\int dw \frac{1}{(w-z)^{2-\Delta}}\ O_{\Delta}(w,\bar{z})
\end{eqnarray}
do have a two-point function similar to those in ordinary 2D CFT \eqref{2ptLight}.
\begin{equation}
   \mathcal{W}(\Delta_{1},\Delta_{2})= \langle L^{+}[\phi_{\Delta_{1}}]L^{-}[\phi_{\Delta_{2}}]\rangle= 2\pi^{3}\delta(\Delta_{1}+\Delta_{2}-2) (z_{2}-z_{1})^{\Delta_{1}-2}(\Bar{z}_{1}-\Bar{z}_{2})^{\Delta_{2}-2}.
\end{equation}
\end{comment}

Another modification of the Mellin basis was introduced in \cite{Banerjee:2018gce}, primarily to regulate the ultra-violet divergences of Mellin-transformed tree-level graviton amplitudes. Interestingly, these modified Mellin operators transform as local fields of definite spin and scaling dimensions on a three-dimensional Carroll manifold (see \cite{Nguyen:2023vfz, Nguyen:2023miw, Bagchi:2023cen} for a recent review of results). Carroll manifolds are null manifolds equipped with geometric structures shared by the null-infinity $\mathcal{I}^{+}$ and isolated horizons alike \cite{Ashtekar:2024mme}. A large number of checks carried out in the past one and a half decades \cite{Barnich:2010eb, bagchi2010correspondence, Barnich:2011mi, bagchi2013holography, detournay2014variational, Bagchi:2016bcd, Jiang:2017ecm, Saha:2022gjw} clearly indicate Carrollian conformal field theories living on the 2 dimensional $\mathcal{I}^{+}$ to be strong candidates as holographic dual to a quantum theory of gravity in 3 dimensional asymptotically flat space-time. More recent explorations \cite{Donnay:2022aba, Donnay:2022wvx, Donnay:2023mrd, Banerjee:2020kaa, Bagchi:2022emh, Mason:2023mti, Saha:2023hsl} provide a compelling set of reasons for considering 3-dimensional conformal Carrollian theories living on $\mathcal{I}^{+}$ to be dual to 4 dimensional asymptotically flat space-time in bulk, at least at the level of tree-level scattering amplitudes of gravitons and gluons. One of the recent observations in this direction was done in \cite{deGioia:2022fcn}, where the flat space limits of AdS Witten diagram computations gave rise to 2D celestial correlation functions. This paved a route parallel to HKLL formulation for getting flat space scattering amplitudes from large radius $(R)$ limit of AdS space. However, this requires smearing the celestial operators on $\mathcal{I}^+$ around the AdS boundary time $\tau = \pm \frac{\pi}{2}$. When one generalizes this to other points $\tau = \pm \frac{\pi}{2} + \frac{u}{R}$, one is expected to recover 3D correlation functions satisfying BMS or Carrollian Ward identities. In this regard, an exhaustive set of checks involving 3-point functions of scalars and gluons were done \cite{Bagchi:2023cen} starting from a large $R$ expansion of AdS Witten diagrams. A similar analysis was done recently in \cite{Alday:2024yyj} where it was shown that the flat limit of AdS Witten diagrams reproduces Carrollian Feynman diagrams \cite{Liu:2024nfc} by identifying the large $R$ limit of AdS bulk with the Carrollian limit of $3D$ flat space-time. For recent developments in this subject, we refer the reader to \cite{deBoer:2021jej, deBoer:2023fnj, Chen:2023pqf, Chen:2024voz, Chen:2021xkw, Ecker:2023uwm, Grumiller:2023rzn, Ciambelli:2023tzb, Aggarwal:2024gfb, Ecker:2024czx, Banerjee:2023jpi, Du:2024tlu, Bagchi:2024efs, Mehra:2024zqv, Mehra:2023rmm}.

It has been established that light-transformed Mellin operators give rise to correlators of ordinary CFTs, hence we were motivated to study the light transformation of modified Mellin ones, particularly keeping in mind the properties of light transformation in the context of finite 
%graviton 
amplitudes. Another motivation is the study of light-transformed operators in Carrollian field theories. As expected, performing a light transformation on the Modified Mellin operator eliminates the ultra-local behaviour of the lower point correlation functions. Consequently, we show that the light-transformed two and three-point functions satisfy all the Poincar\'e Ward identities. To verify the Ward identities, one needs to understand how these symmetry generators act on this new basis, and we observe that the space-time translations act non-trivially on the light-transformed operators. 
\begin{comment}    
As a consequence of this non-triviality, we show that on a light-transformed basis, one can intrinsically obtain the OPE of light-transformed operators \color{red} from a purely symmetry-based analysis without referring to any bulk theory. This is a novel result since, in a generic CFT, one can not fix the leading order singularity using symmetries. We show that the OPE obtained intrinsically is consistent with the collinear limit of the three-point function. \color{black}  While discussing OPE, we also comment on the soft factorization of the light-transformed three-point function and show that it is consistent with the collinear limit of the three-point function. 
\end{comment}
It is well known from \cite{Bagchi:2016bcd} that Ward identities of Carrollian conformal symmetry generators admit solutions that are identical to correlation functions of 2D CFT and we will refer to this branch of solutions as the \textit{2D CFT branch} of Carrollian conformal field theories. These correlation functions are independent of the null time coordinate $u$. Not long ago, in \cite{Bagchi:2022emh}, it was shown that these Ward identities also admit solutions that are ultra-local, and those solutions may be referred to as the \textit{Delta function branch}. Interestingly, the correlation functions of this second branch are identical to the ones obtained in the context of Celestial CFT \cite{Banerjee:2018gce}. It is not clear whether a map exists at the level of correlation functions between these two branches of Carroll CFT. In our second set of results, we show that light transformation provides a possible map between these two branches when one considers all the fields to be inserted at the same $u$.

The paper is organized as follows. In section \ref{SectionModifiedMellin} we recall the modified Mellin transformation, discuss it in the context of Celestial CFT, and point out its relevance in studying conformal Carrollian field theories in the context of flat holography. We end this section by discussing the $\mathcal{BMS}_{4}$ generators and the action of translation as weight-shifting operators on Carrollian conformal primaries. Following this we introduce light transformation in section \ref{SectionLightTransform} and discuss the action of global BMS generators a.k.a Poincar\'e generators on the light-transformed fields and end the section by performing a check on the algebra of these generators. In section \ref{2ptScalar} and \ref{3ptScalar} we find the light-transformed two and three-point function of scalars and show that they satisfy the Ward identities of Poincar\'e transformation. In section \ref{Section2ptGluon} and \ref{3ptGluon} we calculate the two and three-point function of gluons. In section \ref{CarrollConnection} we demonstrate how light transformation provides a map between the two branches of Carroll CFT when one considers operators inserted at the same null time. In \ref{SectionOPEgluon} we obtain our main result on the OPE limit of two light-transformed gluon fields. First, we derive the OPE by taking an intrinsic outlook, i.e., we use the symmetries to fix the leading order singularity and fix the OPE coefficient up to a numerical value which is found by studying the soft limit of the three-point function. Subsequently, we show that this OPE is consistent with the collinear limit of the three-point function. We conclude by discussing possible future directions. 

\section{Modified Mellin transformation and Carroll CFT} \label{SectionModifiedMellin}
The modified Mellin transformation was first introduced in \cite{Banerjee:2018gce} and it was shown in \cite{Banerjee:2019prz} that if we use the Modified Mellin prescription to calculate correlation functions on $\mathcal{I}^{+}$, then the four-point graviton correlation function turns out to be non-divergent. Hence, from this point of view, the coordinate $u$ in \eqref{ModifiedMellin} acts as a regulator in Celestial CFT and can be understood as a parameter. The transformation is defined as follows,  
 \begin{equation}
    \phi_{a,h,\Bar{h}}(u,z,\Bar{z})=N(\epsilon,\Delta)\int d\omega \ \omega^{\Delta-1}\ e^{-i\epsilon \omega u} \phi^{bulk}(\omega,z,\Bar{z}) \label{ModifiedMellin}
\end{equation}
where $N(\epsilon,\Delta)$ is the normalization constant and $(h,\Bar{h})$ can be written in terms of the planar spin $s$ and scaling dimension of the field $\Delta$ as $(h,\Bar{h})=(\frac{\Delta+s}{2},\frac{\Delta-s}{2})$ and $\epsilon=\pm 1$ for outgoing/ incoming particles. Originally, $\Delta$ was understood to belong to the principal continuous series, $\Delta=1+i\lambda$ with $\lambda \in \mathbb{R}$. However, as we will see later in this section, considering $\Delta$ as an integer has a natural explanation in terms of Carrollian CFTs as well. For conciseness, we denote the pair of conformal weights of the field by $\mathfrak{h}=(h,\Bar{h})$.  If we are considering gluons, then the index $a$ denotes the helicity of the gluon, and $a=\pm1$ denotes positive/negative helicity. 
%can take values $u,z,\Bar{z}$ where $a=z$ implies helicity is $1$, $a=\Bar{z}$ implies helicity is $-1$ and \color{red} $a=u$ implies the field should be considered as scalar. \color{black} The index $a$ is absent in the case of scalar fields.

The modified Mellin transformation was derived from the ordinary Mellin transformation by performing a time translation along the null time direction $u$ %\footnote{To begin with $u$ is a real parameter however since are performing time translations on $\mathcal{I}^{+}$ with $u$ we are calling $u$ the null time coordinate.}
and it was shown in \cite{Banerjee:2020kaa} that these operators form a representation of the extended $\mathcal{BMS}_{4}$ group. Very recently, in \cite{Liu:2024nfc}, this transformation was understood from the perspective of asymptotic expansion of a bulk field in flat spacetime. It was shown that modified Mellin transformation shows up naturally while calculating bulk-to-boundary propagator in the \textit{generalized} $(\phi^{bulk})^{4}$ theory\footnote{It was also shown that if one considers the massless $(\phi^{bulk})^{4}$ and tries to calculate the bulk-to-boundary propagator, then one ends up with the Kirchhoff-d’Adh\'{e}mar formula \eqref{Kirchoffdadhemar}.}. This is consistent with Celestial CFT literature where bulk-to-boundary propagators were considered as conformal primary wave functions which could also be obtained by Mellin transformation of bulk momentum modes \cite{Pasterski:2017kqt, Cheung:2016iub}. 

Before we proceed to study properties of light-transformed modified Mellin primaries we must allude to recent literature \cite{Donnay:2022wvx, Donnay:2022aba} which suggests that a Carrollian conformal primary, $\Phi^{\mathcal{I}^{+}}_{a,h,\Bar{h}}$ living on the boundary of our spacetime, $\mathcal{I}^{+}$, is related to the bulk field $\phi^{bulk}(\omega,z,\Bar{z})$ by the Kirchhoff-d’Adh\'{e}mar formula \cite{Penrose:1985bww, Penrose:1980yx}(which is a Fourier transformation in $\omega$), 
\begin{equation}
    \Phi^{\mathcal{I}^{+}}_{a,h,\Bar{h}}(u,z,\Bar{z})=\int d\omega\ e^{-i\epsilon \omega u} \phi^{bulk}(\omega,z,\Bar{z}) \label{Kirchoffdadhemar}.
\end{equation}
One can immediately see that the modified Mellin transformation \eqref{ModifiedMellin} can be written as a $\partial_{u}$-\textit{descendent} of the Carrollian conformal primary $\Phi^{\mathcal{I}^{+}}_{a,h,\Bar{h}}(u,z,\Bar{z})$ (up to a numerical factors) \cite{Mason:2023mti}, 
\begin{equation}
  \partial^{n-1}_{u}\Phi^{\mathcal{I}^{+}}_{a,h,\Bar{h}}(u,z,\Bar{z})\ \overset{n=\Delta}{=}\ (-i\epsilon)^{\Delta-1}\int d\omega \ \omega^{\Delta-1}\ e^{-i\epsilon \omega u} \phi^{bulk}(\omega,z,\Bar{z})= (-i\epsilon)^{\Delta-1} \phi_{a,h,\Bar{h}}(u,z,\Bar{z}) \label{udescendent}
\end{equation}
where $n$ can be naturally understood to be an integer or it can also be understood as being analytically continued to the complex plane so that $n=\Delta=1+i\lambda$ with $\lambda\in \mathbb{R}$. Although, as we will see in the next section, our definition of light transformation works for both these seemingly equivalent prescriptions, we work with \eqref{ModifiedMellin} since we are interested in understanding the implications of light transformation from the perspective of Celestial CFT.\footnote{We must mention that there exists a map between the boundary fields defined in \eqref{Kirchoffdadhemar} and Mellin primaries of \cite{Pasterski:2017kqt} and it would be interesting to understand Light transforms from this perspective as well since the prescription of \cite{Donnay:2022wvx} requires one to integrate over the null time $u$ to arrive at a Celestial conformal primary}.

The action of the extended $\mathcal{BMS}_{4}$ generators \cite{Barnich:2009se, Bagchi:2016bcd} on a field with conformal weight $\mathfrak{h}$ are given by,
\begin{align}
    & \delta_{M_{r,s}}\phi_{a,\mathfrak{h}}(u,z,\Bar{z})=\ z^{r}\Bar{z}^{s}\partial_{u}\phi_{a,\mathfrak{h}}(u,z,\Bar{z}) \nn \\
    & \delta_{L_{n}}\phi_{a,\mathfrak{h}}(u,z,\Bar{z})=\big[z^{n+1}\partial_{z}+(n+1)(h+\frac{u}{2}\partial_{u})z^{n} \big]\phi_{a,\mathfrak{h}}(u,z,\Bar{z}) \nn \\
    & \delta_{\Bar{L}_{n}}\phi_{a,\mathfrak{h}}(u,z,\Bar{z})=\big[\Bar{z}^{n+1}\partial_{\Bar{z}}+(n+1)(\Bar{h}+\frac{u}{2}\partial_{u})\Bar{z}^{n} \big]\phi_{a,\mathfrak{h}}(u,z,\Bar{z}) \label{BMS}.
\end{align}
We will not work with the full infinite dimensional extended $\mathcal{BMS}_{4}$ group and instead will focus on the global part of it. This boils down to  $r,s=0,1$, for which, $M_{r,s}$ generate translations along $\mathcal{I}^{+}$, and $n=0,\pm1$ for $L_n, \bar{L}_n$ which generate global conformal transformation on the Celestial sphere. This is the Poincar\'e group in four dimensions which is equivalent to the global conformal Carrollian group in three dimensions.

 Lastly, we would like to use the fact that the action of $M_{00}$ on the field $\phi_{a,\mathfrak{h}}(u,z,\Bar{z})$ is given by the following variation \eqref{BMS},
\begin{equation}
\delta_{M_{00}}\phi_{a,\mathfrak{h}}(u,z,\Bar{z})=-i\partial_{u}\phi_{a,\mathfrak{h}}(u,z,\Bar{z}) \label{M00}
\end{equation}
Using the definition of the modified Mellin transformation given in equation \eqref{ModifiedMellin} we can rewrite equation \eqref{M00} as 
\begin{equation}
    \delta_{M_{00}}\phi_{a,\mathfrak{h}}(u,z,\Bar{z})=-\epsilon \phi_{a,\mathfrak{h}+\frac{1}{2}}(u,z,\Bar{z}). \label{M00Mellin}
\end{equation}
where $\mathfrak{h}+\frac{1}{2}=(h+\frac{1}{2},\Bar{h}+\frac{1}{2})$ are the conformal weights of the weight shifted conformal primary which is why they are referred to as $\partial_{u}$-\textit{descendant}. This shows that translations act as weight-shifting operators on modified Mellin primaries \cite{Banerjee:2020kaa, Fotopoulos:2019vac}\footnote{The action of translations, as weight-shifting operators were found in \cite{Fotopoulos:2019vac} in the context of ordinary Mellin operators.} and this is consistent with the fact that in \eqref{BMS} we have $[\delta_{L_{0}},\delta_{M_{00}}]=\frac{1}{2}\delta_{M_{00}}$. In the next section, we will use \eqref{BMS} to calculate the action of these generators on the fields defined in \eqref{L+} and \eqref{L-}.

\section{Light transformation}\label{SectionLightTransform}
We define the two types of light transformations (indexed with a `$+$' or `$-$') of the conformal primary as a contour integral on the Celestial sphere as \cite{Banerjee:2022hgc}
\begin{align}
    & L^{+}[\phi_{a,\mathfrak{h}}](u,z,\Bar{z})=\oint_{w=z} dw \ \frac{\phi_{a,\mathfrak{h}}(u,w,\Bar{z})}{(w-z)^{2-2h}} \label{L+} \\
    & L^{-}[\phi_{a,\mathfrak{h}}](u,z,\Bar{z})=\oint_{\Bar{w}=\Bar{z}} d\Bar{w} \ \frac{\phi_{a,\mathfrak{h}}(u,z,\Bar{w})}{(\Bar{w}-\Bar{z})^{2-2\Bar{h}}}\label{L-}
\end{align}
 where we have also assumed that the fields are analytic functions of the coordinates on $\mathcal{I}^{+}$ and that the only singularity in the integrand is at the point $z$ for $L^{+}$ integral and at $\Bar{z}$ for $L^{-}$ integral \footnote{Although massless Celestial conformal primaries have a pole at $q(z,\Bar{z})\cdot X$ where $q_{\mu}(z,\Bar{z})$ is the null momenta and $X^{\mu}$ is a bulk spacetime point, we assume that we can smoothly deform our contour on the Celestial sphere such that it only encloses the poles coming from the denominator of \eqref{L+} and \eqref{L-}.}. Following the convention of \cite{Sharma:2021gcz} we use $L^{+}$ for positive helicity fields and $L^{-}$ for negative helicity fields. This choice was further motivated by \cite{Arkani-Hamed:2009hub}. We do not comment on the conformal weights of the light-transformed operators in this section since, unlike \cite{Kravchuk:2018htv}, we define them as contour integrals. We will fix the weights of the light-transformed primaries by looking at how they transform under scaling $\delta_{L_{0}}$ and $\delta_{\bar{L}_{0}}$ \eqref{L+L0}, \eqref{L-L0}, \eqref{L+Lbar0}, \eqref{L-Lbar0}. \label{BMSLight} Note that while defining the light transformation we have also assumed that $z$ and $\Bar{z}$ are independent coordinates; however, in the end, the light-transformed amplitudes will be defined on the \textit{real surface} \cite{DiFrancesco:1997nk}.

The main objective behind studying light-transformed operators was \cite{Banerjee:2022hgc} to rewrite celestial correlation functions in a basis where they follow the usual power law behaviour of 2D CFT. While a similar motivation of recovering a power law behaviour (or that the light transforms reproduce this power law behaviour in some limit as we will see in section \ref{2ptLight}) still holds for modified Mellin/ Carrollian correlators, one has to approach this carefully by making sure that the resultant, non-distributional, correlation function satisfy the symmetries of the theory. This amounts to studying Ward identities of correlation functions constructed out of light-transformed operators and it is evident from the definition of light transform \eqref{L+},\eqref{L-} that derivatives of $z$ act non-trivially on them. In this section, we will study how the symmetry generators mentioned in \eqref{BMS} act on light-transformed fields and we will see that even though we do not integrate over the coordinate $u$, derivatives of light-transformed fields along the null-time direction are non-trivial. We will use the residue theorem to simplify some of the contour integrals as follows, 
\begin{equation}
 L^{+}[\phi_{a,\mathfrak{h}}](u,z,\Bar{z})=\oint dw \ \frac{\phi_{a,\mathfrak{h}}(u,w,\Bar{z})}{(w-z)^{2-2h}} =\frac{2\pi i}{(1-2h)!}\lim_{w\rightarrow z}(\partial_{w})^{1-2h}\phi_{a,\mathfrak{h}}(u,z,\Bar{z}) \label{residuetheorem}
\end{equation}
 where we have assumed that $2-2h$ is a positive integer and that we can analytically continue to complex values of $\Delta$. Hence, we propose that this prescription is well suited to study light transformation of both kinds of primaries, one obtained from \eqref{Kirchoffdadhemar} and the other from \eqref{ModifiedMellin} although, as we mentioned before, the latter is a \textit{$\partial_{u}$-descendant} of the former.

% We emphasize again that although our notation is built for spin 1 particles it easily simplifies for spin 0 particles as well. 
\subsection {Action of translations on light-transformed operators} \label{PoincareLight}

We start by studying the action of translations on the light-transformed fields. We would like to note that the action of the translation generators on the light-transformed Mellin operators for $\Delta =1$ \eqref{Kirchoffdadhemar} are trivial. In the following, we will keep $\Delta$ arbitrary.
\section*{Action of \texorpdfstring{$M_{00}$, $M_{10}$, $M_{01}$, $M_{11}$}{\265}}
We start with $M_{00}$ and use \eqref{M00Mellin}, 
\begin{align}
    \delta_{M_{00}}L^{+}[\phi_{a,\mathfrak{h}}(u,z,\Bar{z})]&=\oint dw \ \frac{\delta_{M_{00}}\phi_{a,\mathfrak{h}}(u,w,\Bar{z})}{(w-z)^{2-2h}}\nn \\
    &=-i\epsilon  \oint dw \ \frac{\phi_{a,\mathfrak{h+\frac{1}{2}}}(u,w,\Bar{z})}{(w-z)^{2-2h}}= -\frac{i\epsilon}{1-2h} \partial_{z}\oint dw \ \frac{\phi_{a,\mathfrak{h}+\frac{1}{2}}(u,w,\Bar{z})}{(w-z)^{1-2h}}  \label{M00Light}
\end{align}

The last term is nothing but the derivative of $L^{+}[\phi_{a,\mathfrak{h}+\frac{1}{2}}]$, so one can write, 
\begin{equation}
    \delta_{M_{00}}L^{+}[\phi_{a,\mathfrak{h}}]= -\frac{i\epsilon}{1-2h} \partial_{z}L^{+}[\phi_{a,\mathfrak{h}+\frac{1}{2}}] \label{L+M00}
\end{equation}
where we have suppressed the dependence of $\phi_{a,\mathfrak{h}}$ on the Bondi coordinates $(u,z,\Bar{z})$. The appearance of $\partial_{z}$ which, although, might seem non-trivial from a Carrollian or Celestial perspective, naturally arises for light-transformed operators \cite{Banerjee:2022hgc}.  This, however, calls for a better understanding of the representation of light-transformed operators %since such operators are not conformal primaries from a Carrollian perspective \cite{Banerjee:2020vnt, Nguyen:2023vfz}
(we would urge the reader to look at Section 2 of \cite{Saha:2023hsl} for a first principle analysis of the representation of Carrollian fields). We analyze the transformation of $\delta_{M_{00}}L^{+}[\phi_{a,\mathfrak{h}}]$ under $\delta_{L_{0}}$ and $\delta_{L_{1}}$, in Appendix \ref{M00action}, and verify that it is a descendant with conformal weights $(\frac{3}{2}-h,\Bar{h})$. This, along with the result of \eqref{L+L0}, implies that $\delta_{M_{00}}$ acts as a raising operator for light-transformed primaries.

Since the finite transformation corresponding to $\delta_{M_{00}}$ shifts the coordinate $u$ by a constant factor the action of this variation on $L^{-}[\phi_{a,\mathfrak{h}+\frac{1}{2}}]$ is given by a similar expression with $z$ replaced by $\Bar{z}$ and $h$ replaced by $\Bar{h}$ in \eqref{L+M00},
\begin{equation}
     \delta_{M_{00}}L^{-}[\phi_{a,\mathfrak{h}}]= -\frac{i\epsilon}{1-2\Bar{h}} \partial_{\Bar{z}}L^{-}[\phi_{a,\mathfrak{h}+\frac{1}{2}}]. \label{L-M00}
\end{equation}
\begin{comment}
\color{red} [This is a contradiction to the fact that action of $\delta_{M_{00}}$ gives a descendant, we have to decide whether to mention this or not]Note the non-trivial action of the weight shifting operators on light-transformed fields. If we stop at the first equality in \eqref{M00Light}, then we can express the action of $\delta_{M_{00}}$ as a $u$-derivative of $L^{+}[\phi_{a,\mathfrak{h}}]$
\begin{equation}
    \delta_{M_{00}}L^{+}[\phi_{a,\mathfrak{h}}(u,z,\Bar{z})]= \oint dw \ \frac{\delta_{M_{00}}\phi_{a,\mathfrak{h}}(u,w,\Bar{z})}{(w-z)^{2-2h}}=\partial_{u}\oint dw \ \frac{\phi_{a,\mathfrak{h}}(u,w,\Bar{z})}{(w-z)^{2-2h}}=\partial_{u}L^{+}[\phi_{a,\mathfrak{h}}] \label{M00uderivative}.
\end{equation} 
Now, contrary to \eqref{L+M00}, there is no reason to believe that the right-hand side is a descendant of a light-transformed primary. All the more something non-trivial has happened, a derivative along the u-direction has been re-expressed as a derivative in $z/\Bar{z}$, we leave this problem for future works.\color{black}
\end{comment}
Next, we study the action of $\delta_{M_{10}}$ on $L^{+}[\phi_{a,\mathfrak{h}}]$ and note that they generate coordinate dependent translation for the field $\phi_{a,\mathfrak{h}}$,
$$
\delta_{M_{10}}L^{+}[\phi_{a,\mathfrak{h}}]=\oint dw\ \frac{\delta_{M_{10}}\phi_{a,\mathfrak{h}}}{(w-z)^{2-2h}}= -i\epsilon \oint dw \ {\frac{w\phi_{a,\mathfrak{h}+\frac{1}{2}}}{(w-z)^{2-2h}}}.
$$
We employ the residue theorem \eqref{residuetheorem} to get,
$$
\oint dw \ {\frac{w\ \phi_{a,\mathfrak{h}+\frac{1}{2}}}{(w-z)^{2-2h}}}= \frac{2\pi i}{(1-2h)!}\lim_{w\rightarrow z} (\partial_{w})^{1-2h}(w\ \phi_{a,\mathfrak{h}+\frac{1}{2}})
$$
This gives us,
\begin{equation}
     \delta_{M_{10}}L^{+}[\phi_{a,\mathfrak{h}}]=-i\epsilon \Big( \frac{z}{1-2h}\partial_{z}+1 \Big)L^{+}[\phi_{a,\mathfrak{h}+\frac{1}{2}}]. \label{L+M10}
\end{equation}
 However, the action of $\delta_{M_{10}}$ on $L^{-}[\phi_{\Delta}]$ is straightforward since the variable of integration is $\Bar{w}$,
 \begin{equation}
     \delta_{M_{10}}L^{-}[\phi_{a,\mathfrak{h}}]=-\frac{i\epsilon z}{1-2\Bar{h}}\partial_{\Bar{z}}L^{-}[\phi_{a,\mathfrak{h}+\frac{1}{2}}]. \label{L-M10}
 \end{equation}
Similar analysis for $\delta_{M_{01}}$ and $\delta_{M_{11}}$ give us, 
\begin{align}
& \delta_{M_{01}}L^{+}[\phi_{a,\mathfrak{h}}]=-\frac{i\epsilon \Bar{z}}{1-2h}\partial_{z}L^{+}[\phi_{a,\mathfrak{h}+\frac{1}{2}}] \label{L+M01} \\ 
    & \delta_{M_{01}}L^{-}[\phi_{a,\mathfrak{h}}]=-i\epsilon \Big( \frac{\Bar{z}}{1-2\Bar{h}}\partial_{\Bar{z}}+1 \Big)L^{-}[\phi_{a,\mathfrak{h}+\frac{1}{2}}] \label{L-M01}\\
    & \delta_{M_{11}}L^{+}[\phi_{a,\mathfrak{h}}]=-i\epsilon \Bar{z}\Big(\frac{z}{1-2h}\partial_{z}+ 1\Big)L^{+}[\phi_{a,\mathfrak{h}+\frac{1}{2}}] \label{L+M11}\\
    & \delta_{M_{11}}L^{-}[\phi_{a,\mathfrak{h}}]=-i\epsilon z \Big(\frac{\Bar{z}}{1-2\Bar{h}}\partial_{\Bar{z}}+1\Big)L^{-}[\phi_{a,\mathfrak{h}+\frac{1}{2}}] . \label{L-M11}
\end{align}
Note that the action of these generators on the light-transformed fields is non-trivial in contrast to the action on the celestial conformal primaries given in \eqref{BMS}. Apart from this, it is easy to check that by taking linear combinations of these generators, given above, we can reproduce the action of $4D$ bulk translations on fields obtained by performing a light transformation of the ordinary Mellin basis \cite{Banerjee:2022hgc}. This is well known in Celestial CFT literature from \cite{Banerjee:2020kaa, Stieberger:2018onx, Fotopoulos:2019vac}.

\section*{Action of Lorentz transformations}
Lorentz transformations act as the group of global conformal transformation on the Celestial sphere and they consist of 2D-``translation" (on the Celestial sphere), scaling and special conformal transformation. We start with translations on Celestial sphere and recall that, in the context of the highest weight representation of a conformal field theory, it is the translations that act as raising operators on the highest weight states and gives us \textit{descendants}\footnote{Look at \cite{Delgado:2021hmf} for a nice review of the highest weight representation for CFTs in $d>2$}. We now look at the action of the generators of translation on the light-transformed fields, 
$$
 \delta_{L_{-1}}L^{+}[\phi_{a,\mathfrak{h}}]=\oint dw\ \frac{\partial_{w}\phi_{a,\mathfrak{h}}}{(w-z)^{2-2h}} 
$$
On simplification, residue theorem gives us, 
\begin{equation}
     \delta_{L_{-1}}L^{+}[\phi_{a,\mathfrak{h}}]=\partial_{z}L^{+}[\phi_{a,\mathfrak{h}}] \label{L+L-1}
\end{equation}
similarly, 
\begin{equation}
     \delta_{L_{-1}}L^{-}[\phi_{a,\mathfrak{h}}]=\partial_{z}L^{-}[\phi_{a,\mathfrak{h}}]. \label{L-L-1}
\end{equation}
   The action of $\delta_{\Bar{L}_{-1}}$ on the light-transformed fields are,
   \begin{align}
       & \delta_{\Bar{L}_{-1}}L^{+}[\phi_{a,\mathfrak{h}}]=\partial_{\Bar{z}}L^{+}[{\phi_{a,\mathfrak{h}}}] \label{L+Lbar-1}\\
       &\delta_{\Bar{L}_{-1}}L^{-}[\phi_{a,\mathfrak{h}}]=\partial_{\Bar{z}}L^{-}[{\phi_{a,\mathfrak{h}}}] .\label{L-Lbar-1}
   \end{align}
Now, as we mentioned before, the understanding of translation from the perspective of Celestial CFT is that they are realized as weight-shifting operators on the Celestial sphere \cite{Stieberger:2018onx}. However, from the perspective of a generic CFT, one would expect any translation on the conformal sphere to be realised as operations that give us descendant fields. Comparing the above expressions for ``translations" on the Celestial sphere with equations \eqref{L+M00} and \eqref{L-M00}, we see that, in the light-transformed basis, bulk spacetime translations can be realised as translations on the celestial sphere.  

Next, we turn to the scaling operator $L_{0}$ and $\Bar{L}_{0}$ and find that
\begin{comment}
\begin{align}
   & \delta_{L_{0}}L^{+}[\phi_{a,\mathfrak{h}}]=-i\Big(z\partial_{z}L^{+}[\phi_{a,\mathfrak{h}}]+(1-2h)L^{+}[\phi_{\Delta}]+hL^{+}[\phi_{\Delta}] -\frac{i \epsilon u }{2(1-2h)}\partial_{z}L^{+}[\phi_{a,\mathfrak{h}+\frac{1}{2}}] \Big) \label{L+L0} \\
 & \delta_{L_{0}}L^{-}[\phi_{\Delta}]=-i\Big(z\partial_{z}L^{-}[\phi_{a,\mathfrak{h}}]+hL^{-}[\phi_{a,\mathfrak{h}}] -\frac{i \epsilon u }{2(1-2\Bar{h})}\partial_{\Bar{z}}L^{-}[\phi_{a,\mathfrak{h}+\frac{1}{2}}] \Big). \label{L-L0}   
\end{align}    
\end{comment}
\begin{align}
   & \delta_{L_{0}}L^{+}[\phi_{a,\mathfrak{h}}]=\Big(z\partial_{z}L^{+}[\phi_{a,\mathfrak{h}}]+(1-h)L^{+}[\phi_{\Delta}]-\frac{i \epsilon u }{2(1-2h)}\partial_{z}L^{+}[\phi_{a,\mathfrak{h}+\frac{1}{2}}] \Big) \label{L+L0} \\
 & \delta_{L_{0}}L^{-}[\phi_{\Delta}]=\Big(z\partial_{z}L^{-}[\phi_{a,\mathfrak{h}}]+hL^{-}[\phi_{a,\mathfrak{h}}] -\frac{i \epsilon u }{2(1-2\Bar{h})}\partial_{\Bar{z}}L^{-}[\phi_{a,\mathfrak{h}+\frac{1}{2}}] \Big). \label{L-L0}   
\end{align}
By similar analysis we find, 
\begin{comment}
    \begin{align}
    & \delta_{\Bar{L}_{0}}L^{+}[\phi_{a,\mathfrak{h}}]=-i\Big(\Bar{z}\partial_{\Bar{z}}L^{+}[\phi_{a,\mathfrak{h}}]+\Bar{h}L^{+}[\phi_{a,\mathfrak{h}}] -\frac{i \epsilon u }{2(1-2h)}\partial_{{z}}L^{+}[\phi_{a,\mathfrak{h}+\frac{1}{2}}] \Big) \label{L+Lbar0} \\
    & \delta_{\Bar{L}_{0}}L^{-}[\phi_{a,\mathfrak{h}}]=-i\Big(\Bar{z}\partial_{\Bar{z}}L^{-}[\phi_{a,\mathfrak{h}}]+(1-2\Bar{h})L^{-}[\phi_{a,\mathfrak{h}}]+\Bar{h}L^{-}[\phi_{a,\mathfrak{h}}] -\frac{i \epsilon u }{2(1-2\Bar{h})}\partial_{\Bar{z}}L^{-}[\phi_{a,\mathfrak{h}+\frac{1}{2}}] \Big).  \label{L-Lbar0}
\end{align}
\end{comment}

\begin{align}
    & \delta_{\Bar{L}_{0}}L^{+}[\phi_{a,\mathfrak{h}}]=\Big(\Bar{z}\partial_{\Bar{z}}L^{+}[\phi_{a,\mathfrak{h}}]+\Bar{h}L^{+}[\phi_{a,\mathfrak{h}}] -\frac{i \epsilon u }{2(1-2h)}\partial_{{z}}L^{+}[\phi_{a,\mathfrak{h}+\frac{1}{2}}] \Big) \label{L+Lbar0} \\
    & \delta_{\Bar{L}_{0}}L^{-}[\phi_{a,\mathfrak{h}}]=\Big(\Bar{z}\partial_{\Bar{z}}L^{-}[\phi_{a,\mathfrak{h}}]+(1-\Bar{h})L^{-}[\phi_{a,\mathfrak{h}}] -\frac{i \epsilon u }{2(1-2\Bar{h})}\partial_{\Bar{z}}L^{-}[\phi_{a,\mathfrak{h}+\frac{1}{2}}] \Big).  \label{L-Lbar0}
\end{align}
This tells us that if we perform a light transformation $L^{+}$ on $\phi_{a,\mathfrak{h}}$ the resulting primary has weights $\tilde{\mathfrak{h}}=(1-h,\Bar{h})$ \eqref{L+L0} and if we perform $L^{-}$ on $\phi_{a,\mathfrak{h}}$ we get a primary with weights $\tilde{\mathfrak{h}}=(h,1-\Bar{h})$\eqref{L-L0}.
Lastly, we look at special conformal transformations and find that, 
\begin{align}
    & \delta_{L_{1}}L^{+}[\phi_{a,\mathfrak{h}}]=\Bigg[ z^{2}\partial_{z}L^{+}[\phi_{a,\mathfrak{h}}]+2(1-h)z L^{+}[\phi_{a,\mathfrak{h}}] -i \epsilon u \bigg(\frac{z}{(1-2h)}\partial_{z}+1\bigg)L^{+}[\phi_{a,\mathfrak{h}+\frac{1}{2}}]  \Bigg] \label{L+L1} \\
    & \delta_{L_{1}}L^{-}[\phi_{a,\mathfrak{h}}]=  \Big[ z^{2}\partial_{z}L^{-}[\phi_{a,\mathfrak{h}}]+2h z L^{-}[\phi_{a,\mathfrak{h}}]-\frac{i \epsilon u}{(1-2\Bar{h})}z \partial_{\Bar{z}}L^{-}[\phi_{a,\mathfrak{h}+\frac{1}{2}}] \Big] \label{L-L1}.
\end{align}
Similarly,
\begin{align}
    & \delta_{\Bar{L}_{1}}L^{+}[\phi_{a,\mathfrak{h}}]=  \Big[ \Bar{z}^{2}\partial_{\Bar{z}}L^{+}[\phi_{a,\mathfrak{h}}]+ 2\Bar{h} \Bar{z} L^{+}[\phi_{a,\mathfrak{h}}]-\frac{i \epsilon u}{(1-2h)} \Bar{z} \partial_{{z}}L^{+}[\phi_{a,\mathfrak{h}+\frac{1}{2}}] \Big] \label{L+Lbar1} \\
    & \delta_{\Bar{L}_{1}}L^{-}[\phi_{a,\mathfrak{h}}]=\Bigg[ \Bar{z}^{2}\partial_{\Bar{z}}L^{-}[\phi_{a,\mathfrak{h}}]+2(1-\Bar{h})\Bar{z} L^{-}[\phi_{a,\mathfrak{h}}] -i \epsilon u \bigg(\frac{\Bar{z}}{(1-2\Bar{h})}\partial_{\Bar{z}}+ 1\bigg)L^{-}[\phi_{a,\mathfrak{h}+\frac{1}{2}}].  \Bigg] \label{L-Lbar1}
\end{align}
One can conclude from the above relations that the action of Lorentz transformations on light-transformed field $L^{+}[\phi_{a,\mathfrak{h}}]$ is similar to a conformal primary with weight $(1-h,\bar{h})$ and action of Lorentz transformations on light-transformed field $L^{-}[\phi_{a,\mathfrak{h}}]$ is similar to a conformal primary with weight $(h,1-\bar{h})$. Thus, light-transformed operators are primaries from the viewpoint of Celestial CFT. 
%but not from the viewpoint of Carrollian CFT owing to \eqref{L+M00} and \eqref{L-M00} as we show in Appendix \ref{M00action}. 
%\textcolor{red}{As we mentioned before, we study how these transformations act on fields given in \eqref{L+M00} and \eqref{L-M00} in Appendix \ref{M00action}.}

%\section{S-Algebra and spectral flow}
%In section \ref{BMSLight} we saw that a derivative along $u$ could be re-written as a derivative along $z$. 
\section{Light transformed correlators: Scalars and Gluons}
\subsection{Two point function of Scalars}\label{2ptScalar}
In this section, we study scalar fields and note that for scalars $h=\Bar{h}=\frac{\Delta}{2}$. Hence the spin label is omitted in this section. The modified Mellin transformed two-point function is given by \cite{Banerjee:2018gce, Bagchi:2022emh},
\begin{equation}
    \langle \phi_{\Delta_{1}} \phi_{\Delta_{2}}\rangle= 4\pi^{2}\frac{\Gamma(\Delta_{1}+\Delta_{2}-2)}{i^{\Delta_{1}+\Delta_{2}-2}} \frac{\delta^{2}(z_{1}-z_{2})}{(u_{1}-u_{2})^{\Delta_{1}+\Delta_{2}-2}}. \label{2ptMellin}
\end{equation}
wherein we used $\epsilon_{1}=-1$ and $\epsilon_{2}=1$ in \eqref{ModifiedMellin}. This is the \textit{Delta function branch} of Carrollian CFT. Performing a light transformation using \eqref{L-} and \eqref{L+} we get,
\begin{equation}
    \langle L^{+}[\phi_{\Delta_{1}}]L^{-}[\phi_{\Delta_{2}}]\rangle=4\pi^{3}\frac{\Gamma(\Delta_{1}+\Delta_{2}-2)}{(i(u_{1}-u_{2}))^{\Delta_{1}+\Delta_{2}-2}} \oint_{w_{1}=z_{1}} dw_{1}\oint_{\Bar{w}_{2}=\Bar{z}_{2}} d\Bar{w}_{2} \frac{\delta(w_{1}-z_{2})\delta(\Bar{z}_{1}-\Bar{w}_{2})}{(w_{1}-z_{1})^{2-\Delta_{1}}(\Bar{w}_{2}-\Bar{z}_{2})^{2-\Delta_{2}}}
\end{equation}
where the Dirac delta function has a complex argument and can be understood as
\begin{equation}
\delta(x)=\lim_{\nu \rightarrow 0}\frac{\Gamma(x)}{\nu^{x}}\quad; x\in \mathbb{C} \quad \text{and} \quad \nu >0. \label{Diracdelta}
\end{equation}
This representation of the Dirac delta function was used in \cite{Donnay:2020guq} in the context of Mellin transformation and in \cite{Banerjee:2022hgc} in the context of light transformation. Performing the integrals by deforming the contour appropriately to enclose the poles of the Gamma function we get,
\begin{equation}
   \mathcal{W}(\Delta_{1},\Delta_{2})= \langle L^{+}[\phi_{\Delta_{1}}]L^{-}[\phi_{\Delta_{2}}]\rangle= 4\pi^{3}\frac{\Gamma(\Delta_{1}+\Delta_{2}-2)}{i^{\Delta_{1}+\Delta_{2}-2}} \frac{(z_{2}-z_{1})^{\Delta_{1}-2}(\Bar{z}_{1}-\Bar{z}_{2})^{\Delta_{2}-2}}{(u_{1}-u_{2})^{\Delta_{1}+\Delta_{2}-2}}. \label{2ptLight}
\end{equation}
This two-point function has a resemblance with the usual 2D CFT two-point function since we have recovered the characteristic power law behaviour. However, we must note that this two-point function is non-zero even when the conformal weights of the two operators are different. 
%This is reminiscent of the fact that these are derived from correlation functions of Carrollian conformal field theories\cite{Bagchi:2022emh, Donnay:2022wvx}, which are non-relativistic CFTs.

\subsubsection{Ward Identity}\label{2ptWardIdentity}
In this section, we study the Ward Identity corresponding to the Lorentz transformations and translations given in section \ref{BMSLight}. We would like to point out that one could take a bottom-up approach to solve a set of differential equations corresponding to the Ward identities and find a unique solution for the correlators. However, because of the non-trivial weight-shifting action of translations, given in \ref{PoincareLight}, the differential equations become coupled and hence cumbersome. So, we use the correlators computed in \cite{Bagchi:2023cen, Bagchi:2023fbj, Mason:2023mti}, light transform them, and check whether they satisfy the Ward identities corresponding to these transformations.
%We will proceed with calculation by noting the action of the symmetries on the light-transformed correlator, which would be the left-hand side of our differential equation, following which we will write the penultimate step which shows that they satisfy the Ward identity.

\begin{itemize}

\item Ward identity for \boldmath{$\delta_{M_{11}}$}
\unboldmath{}

We have from \eqref{L+M11} and \eqref{L-M11}, 
\begin{align}
  \delta_{M_{11}}\langle L^{+}[\phi_{\Delta_{1}}]L^{-}[\phi_{\Delta_{2}}]\rangle= -i\epsilon_{1}\Bar{z}_{1}\Big(\frac{z_{1}}{1-\Delta_{1}}\partial_{z_{1}} +1 \Big) & \mathcal{W}(\Delta_{1}+1,\Delta_{2}) \nn \\
  &-i\epsilon_{2}z_{2}\Big(\frac{\Bar{z}_{2}}{1-\Delta_{2}}\partial_{\Bar{z}_{2}}+1 \Big)\mathcal{W}(\Delta_{1},\Delta_{2}+1).
\end{align}
Substituting for $\mathcal{W}(\Delta_{1},\Delta_{2})$ from \eqref{2ptLight} we get, 
\begin{equation}
\delta_{M_{11}}\langle L^{+}[\phi_{\Delta_{1}}]L^{-}[\phi_{\Delta_{2}}]\rangle=\frac{\Gamma(\Delta_{1}+\Delta_{2}-1)\Bar{z}^{\Delta_{2}-2}_{12}z^{\Delta_{1}-2}_{21}}{(iu_{12})^{\Delta_{1}+\Delta_{2}-1}}\Bar{z}_{1}z_{2}(i\epsilon_{1}+i\epsilon_{2})=0 
 \label{M11Ward}
\end{equation}
We would like point out that the factor of $\epsilon_{1}+\epsilon_{2}=0$ holds trivially because we started with $\epsilon_{1}=-\epsilon_{2}=-1$ in \eqref{2ptMellin}. This can be understood from a kinematic point of view as the two-point correlation function is physical only when one particle is incoming and the other is outgoing.

\item Ward Identity for $\delta_{L_{1}}$ and $\delta_{\Bar{L}_{1}}$

The Ward Identity for $\delta_{\Bar{L}_{1}}$ give us, 
\begin{align}
     \delta_{\Bar{L}_{1}}&\langle L^{+}[\phi_{\Delta_{1}}]L^{-}[\phi_{\Delta_{2}}]\rangle=\Bigg[ \Big( \sum_{i=1}^{2}\Bar{z}^{2}_{i}\partial_{\Bar{z}_{i}}  +\big(\Delta_{1}\Bar{z}_{1}+(2-\Delta_{2})\Bar{z}_{2} \big) \Big)\mathcal{W}(\Delta_{1},\Delta_{2})\nn \\
     &-\frac{i \epsilon_{1}  u_{2}}{(1-\Delta_{1})}\Bar{z}_{1}\partial_{z_{1}} \mathcal{W}(\Delta_{1}+1,\Delta_{2}) -i\epsilon_{2} u_{2}\bigg(\frac{\Bar{z_{2}}}{(1-\Delta_{2})}\partial_{\Bar{z}_{2}}+ 1 \bigg) \mathcal{W}(\Delta_{1},\Delta_{2}+1)  \Bigg] 
\end{align}
where we will use $\Gamma(\Delta_{1}+\Delta_{2}-1)=(\Delta_{1}+\Delta_{2}-2)\Gamma(\Delta_{1}+\Delta_{2}-2)$ to simplify the terms in which weights have been shifted. We will also use the fact that $\epsilon_{1}=-\epsilon_{2}=1$. We get the following simplified expression,
\begin{equation}
   \delta_{\Bar{L}_{1}}\langle L^{+}[\phi_{\Delta_{1}}]L^{-}[\phi_{\Delta_{2}}]\rangle=\Big[ (\Delta_{2}-2)(\Bar{z}_{1}+\Bar{z}_{2})+(2-\Delta_{2})\Bar{z}_{2}+\Delta_{1}\Bar{z}_{1}-(\Delta_{1}+\Delta_{2}-2)\Bar{z}_{1} \Big]\mathcal{W}(\Delta_{1},\Delta_{2})=0. \label{Lbar1Ward} 
\end{equation}
it is easy to see that the right-hand side sums up to zero showing that the light-transformed two-point function satisfies the Ward Identity for $\delta_{\Bar{L}_{1}} $.

We proceed to calculate the Ward Identity for $\delta_{L_{1}}$ by following similar arguments, 
\begin{align}
     \delta_{L_{1}}\langle L^{+}[\phi_{\Delta_{1}}]&L^{-}[\phi_{\Delta_{2}}]\rangle=\Bigg[\bigg(\sum_{i=1}^{2}z^{2}_{i}\partial_{{z}_{i}}+\Big((2-\Delta_{1})z_{1}+\Delta_{2}z_{2}\Big)\bigg)\mathcal{W}(\Delta_{1},\Delta_{2})\nn \\ 
     &-\epsilon_{1}i u_{1}\bigg(\frac{z_{1}}{(1-\Delta_{1})}\partial_{z_{1}}+1 \bigg)\mathcal{W}(\Delta_{1}+1,\Delta_{2})-\frac{\epsilon_{2}i u_{2}}{(1-\Delta_{2})}\partial_{\Bar{z}_{2}}\mathcal{W}(\Delta_{1},\Delta_{2}+1) \Bigg].
\end{align}
Again, using the same techniques for simplification one can see that the two-point function satisfies the Ward identity for $\delta_{L_{1}}$,
\begin{equation}
    \delta_{L_{1}}\langle L^{+}[\phi_{\Delta_{1}}]L^{-}[\phi_{\Delta_{2}}]\rangle=\Big[(\Delta_{1}-2)(z_{1}+z_{2})+(2-\Delta_{1})z_{1}+\Delta_{2}z_{2}-(\Delta_{1}+\Delta_{2}-2)z_{2} \Big]\mathcal{W}(\Delta_{1},\Delta_{2})=0 .\label{L1Ward}
\end{equation}

\end{itemize}
 Since the calculation of Ward identities are repetitive we show the results for $\delta_{M_{11}}$, $\delta_{L_{1}}$ and $\delta_{\Bar{L}_{1}}$. We defer the calculation of the rest of the Ward identities to Appendix \ref{WardIdentity2ptAppendix} since the other Ward identities can be proved by following similar steps.

\subsection{Scalar three-point correlator in (2,2) signature} \label{3ptScalar}
We start with the Modified Mellin transformed field given in \cite{Bagchi:2023cen},
\begin{align}
    \langle \phi_{\Delta_1}(u_{1},z_{1},\Bar{z}_{1})\phi_{\Delta_2}&(u_{2},z_{2},\Bar{z}_{2})\phi_{\Delta_{3}}(u_{3},z_{3},\Bar{z}_{3}) \rangle \ = \ (-1)^{\Delta_{1}+\Delta_{2}-2}\Gamma(\sum^{3}_{i=1}\Delta_{i}-4)\delta(\Bar{z}_{12})\delta(\Bar{z}_{23})\  \nn \\
& \times \ z^{\Delta_{3}-2}_{12} z^{\Delta_{1}-2}_{23} \ z^{\Delta_{2}-2}_{31} [i(z_{1}u_{32}+z_{2}u_{13}+z_{3}u_{21})]^{4-\sum^{3}_{i=1}\Delta_{i}} \prod_{j=1}^{3}\tau(\sigma^{*}_{j})
\end{align}
where $\prod_{j=1}^{3}\tau(\sigma^{*}_{j})$ is a product of channel-dependent indicator functions which enforces the $z_{i}$'s to be ordered in a certain way. For brevity let us define $\sum^{3}_{i=1}\Delta_{i}=\beta$ and we define $\mathcal{M}(\Delta_{1},\Delta_{2},\Delta_{3})$ as, 
\begin{align}
  \mathcal{M}(\Delta_{1},\Delta_{2},\Delta_{3})= \langle \phi_{\Delta_1}\phi_{\Delta_2}\phi_{\Delta_{3}} \rangle  = & \ 4\pi^{2}(-1)^{\Delta_{1}+\Delta_{2}-2}\Gamma(\beta-4)\delta(\Bar{z}_{12})\delta(\Bar{z}_{23})z^{\Delta_{3}-2}_{12} \nn \\
& \times z^{\Delta_{1}-2}_{23} \ z^{\Delta_{2}-2}_{31} \ [i(z_{1}u_{32}+z_{2}u_{13}+z_{3}u_{21})]^{4-\beta}\prod_{j=1}^{3}\tau(\sigma^{*}_{j}).  \label{3ptScalarMellin}
\end{align}
This channel, as a consequence of the indicator functions \cite{Pasterski:2017ylz, Mizera:2022sln}, has support on $(z_{1}<z_{3}<z_{2})$ or $(z_{2}<z_{3}<z_{1})$ and for $\epsilon_{1}=\epsilon_{2}=-1$ and $\epsilon_{3}=1$. 

We will start by performing two light transformations \eqref{L-} on the first two field $\phi_{\Delta_{1}}$ and $\phi_{\Delta_{2}}$, we get, 
\begin{align}
    \langle L^{-}[\phi_{\Delta_1}]L^{-}[\phi_{\Delta_2}]\phi_{\Delta_{3}} \rangle = & \ 4\pi^{2} (-1)^{\Delta_{1}+\Delta_{2}-2}\Gamma(\beta-4)\Bar{z}^{\Delta_{1}-2}_{31}\Bar{z}^{\Delta_{2}-2}_{32}z^{\Delta_{3}-2}_{12} \nn \\
& \times z^{\Delta_{1}-2}_{23} \ z^{\Delta_{2}-2}_{31} \ [i(z_{1}u_{32}+z_{2}u_{13}+z_{3}u_{21})]^{4-\beta}\prod_{j=1}^{3}\tau(\sigma^{*}_{j}). \label{3pt2Light}
\end{align}
Note that since we are working in (2,2) signature, the integrals given in \eqref{L-} are not contour integrals but are over the real line from $[-\infty,\infty]$. As a next step, we can perform a light transformation on the third field using \eqref{L+}, keeping in mind that the limits of the integral are restricted to interval $[z_{1},z_{2}]$. This gives us, 
\begin{align}
   \mathcal{W}(\Delta_{1},\Delta_{2},\Delta_{3})= \langle L^{-}[\phi_{\Delta_1}]L^{-}[\phi_{\Delta_2}]L^{+}[\phi_{\Delta_{3}}] \rangle  =  &\ 4\pi^{2} (-1)^{\Delta_{1}+\Delta_{2}-2}\Gamma(\beta-4)\Bar{z}^{\Delta_{1}-2}_{31}\Bar{z}^{\Delta_{2}-2}_{32}z^{\Delta_{3}-2}_{12} \nn \\
& \times \mathrm{sgn}(z_{21}) \int_{z_{1}}^{z_{2}}dw_{3}\frac{(z_{2}-w_{3})^{\Delta_{1}-2}\ (w_{3}-z_{1})^{\Delta_{2}-2}}{(w_{3}-z_{3})^{2-\Delta_{3}} \ U^{\beta-4}}. \label{3ptScalarLight}
\end{align}
where the $\mathrm{sgn}(z_{12})$ is reminiscent of the ordering mentioned below \eqref{3ptScalarMellin} and we have defined 
\begin{equation} U=i(z_{1}u_{32}+z_{2}u_{13}+w_{3}u_{21})=i\Big((z_{2}-w_{3})u_{1}+(w_{3}-z_{1})u_{2}+z_{12}u_{3} \Big).
\end{equation}
For the purpose of checking the Ward identities for $\mathcal{W}(\Delta_{1},\Delta_{2},\Delta_{3})$ above, we need not compute the exact integral, since we know how the derivatives act on the functions. We would like to point to the fact that \eqref{3ptScalarMellin} is the correlation function for only one of the three channels of the three-point function, and we choose to work with this channel and study its Ward identity. We give an explanation as to why we make this choice while discussing gluon correlators in Section \ref{3ptGluon}.  %The integral in the above equation is tedious to compute and all the more we can check the global BMS Ward identities for this Light transformed three-point function without performing the integral explicitly since we know how the derivatives act on these functions. 

\subsubsection{Ward Identity}\label{3ptWardIdentity}
We start by recalling from \eqref{3ptScalarMellin}, that we are working with $\epsilon_{1}=\epsilon_{2}=-1$ and $\epsilon_{3}=1$. We follow the analysis of section \ref{2ptWardIdentity} but before we proceed to show our results let us briefly review the Leibniz integral theorem which will be necessary for simplifying our calculations.
\subsubsection*{Leibniz integral theorem}\label{Leibniz theorem}
 Note that when we want to check Ward identities for \eqref{L-Lbar1} or \eqref{L+L1} we have to take the derivative of $\mathcal{W}(\Delta_{1},\Delta_{2},\Delta_{3})$ with respect to $z_{1}$ or $z_{2}$. The action of the derivative is non-trivial owing to the fact the limits of the integral in \eqref{3ptScalarLight} involve $z_{1}$ and $z_{2}$. This non-trivial action is given by the Leibniz integral rule which states that, 
 \begin{equation}
     \partial_{\alpha}\int_{v_{1}(\alpha)}^{v_{2}(\alpha)} f(x,\alpha)=f(x,v_{2})\partial_{\alpha}v_{2}-f(x,v_{1})\partial_{\alpha}v_{1}+\int_{v_{1}(\alpha)}^{v_{2}(\alpha)}\partial_{\alpha}f(x,\alpha) \label{LeibnizRule}
 \end{equation}
provided that the function $f(x,\alpha)$ and $\partial_{x}f(x,\alpha)$ is continuous in $\alpha$ and $x$. Now, if we apply this to the integral in \eqref{3ptScalarLight} we get 
\begin{align}
    \partial_{z_{1}}\int_{z_{1}}^{z_{2}}dw_{3}&\frac{(z_{2}-w_{3})^{\Delta_{1}-2}\ (w_{3}-z_{1})^{\Delta_{2}-2}}{(w_{3}-z_{3})^{2-\Delta_{3}} \ U^{\beta-4}}=-\frac{z_{21}^{\Delta_{1}-2}z_{11}^{\Delta_{2}-2}}{z_{13}^{2-\Delta_{3}}(iz_{12}u_{31})^{\beta-4}} \nn \\
    & +\int_{z_{1}}^{z_{2}}dw_{3}\frac{(z_{2}-w_{3})^{\Delta_{1}-2}}{(w_{3}-z_{3})^{2-\Delta_{3}}}\Bigg((2-\Delta_{2})\frac{(w_{3}-z_{1})^{\Delta_{2}-3}}{U^{\beta-4}}+(4-\beta)\frac{(w_{3}-z_{1})^{\Delta_{2}-2}}{U^{\beta-3}}iu_{32} \Bigg) \label{divergencez1}
\end{align}
 similarly for $\partial_{z_{2}}$ we have   
\begin{align}
\partial_{z_{2}}\int_{z_{1}}^{z_{2}}dw_{3}&\frac{(z_{2}-w_{3})^{\Delta_{1}-2}\ (w_{3}-z_{1})^{\Delta_{2}-2}}{(w_{3}-z_{3})^{2-\Delta_{3}} \ U^{\beta-4}}=\frac{z_{22}^{\Delta_{1}-2}z_{21}^{\Delta_{2}-2}}{z_{23}^{2-\Delta_{3}}(iz_{12}u_{32})^{\beta-4}} \nn \\
    & \int_{z_{1}}^{z_{2}}dw_{3}\frac{(w_{3}-z_{1})^{\Delta_{2}-2}}{(w_{3}-z_{3})^{2-\Delta_{3}}}\Bigg( (\Delta_{1}-2)\frac{(z_{2}-w_{3})^{\Delta_{1}-3}}{U^{\beta-4}}+(4-\beta)\frac{(z_{2}-w_{3})^{\Delta_{1}-2}}{U^{\beta-3}}iu_{13} \Bigg)  \label{divergencez2}
\end{align}
where both the results are divergent because of the $z_{11}$ and $z_{22}$ factors. To cancel such divergent quantities we will rewrite the action of the terms containing $\partial_{z_{3}}$, in the forthcoming sections, appearing in the action of $\delta_{L_{0}}$ and $\delta_{L_{\pm 1}}$ so that they cancel such divergences.

We show the results of the Ward identity calculation for some of the generators, namely $\delta_{M_{11}}$, $\delta_{L_{-1}}$ and $\delta_{\Bar{L}_{-1}}$, since the calculations for the other Ward identities are alike to the ones given here. We refer the reader to Appendix \ref{WardIdentity3ptAppendix} for the rest of the calculations.
\begin{itemize}

 \item Ward Identity for \boldmath{$\delta_{M_{11}}$} \unboldmath{}

 From equations \eqref{L+M11} and \eqref{L-M11} we get, 
 \begin{align}
     \delta_{M_{11}}&\langle L^{-}[\phi_{\Delta_1}]L^{-}[\phi_{\Delta_2}]L^{+}[\phi_{\Delta_{3}}] \rangle=-i\epsilon_{1}z_{1}\Big(\frac{\Bar{z}_{1}}{1-\Delta_{1}}\partial_{\Bar{z}_{1}}+1 \Big)\mathcal{W}(\Delta_{1}+1,\Delta_{2},\Delta_{3})\nn \\
     & -i\epsilon_{2}z_{2}\Big( \frac{\Bar{z}_{2}}{1-\Delta_{2}}\partial_{\Bar{z}_{2}}+1  \Big)\mathcal{W}(\Delta_{1},\Delta_{2}+1,\Delta_{3})-i\epsilon_{3}\Bar{z}_{3}\Big(\frac{z_{3}}{1-\Delta_{3}}\partial_{z_{3}}+1 \Big)\mathcal{W}(\Delta_{1},\Delta_{2},\Delta_{3}+1).
 \end{align}
Implementing \eqref{3ptScalarLight} we get,
\begin{align}
      &\delta_{M_{11}}\langle L^{-}[\phi_{\Delta_1}]L^{-}[\phi_{\Delta_2}]L^{+}[\phi_{\Delta_{3}}] \rangle=4\pi^{2}\Gamma(\beta-3)(-1)^{\Delta_{1}+\Delta_{2}-2}\Bar{z}^{\Delta_{1}-2}_{31}\Bar{z}^{\Delta_{2}-2}_{32}\int_{z_{1}}^{z_{2}}dw_{3}\ \Bigg[ z^{\Delta_{3}-2}_{12}\nn \\
    & \times \frac{(z_{2}-w_{3})^{\Delta_{1}-2}\ (w_{3}-z_{1})^{\Delta_{2}-2}}{(w_{3}-z_{3})^{2-\Delta_{3}} \ U^{\beta-3}}i[z_{1}z_{2}\Bar{z}_{3}(\epsilon_{1}-\epsilon_{2})+\Bar{z}_{3}w_{3}\big(-z_{1}(\epsilon_{1}+\epsilon_{3})+z_{2}(\epsilon_{2}+\epsilon_{3}) \big)]\Bigg]=0 \label{3ptM11Ward}
\end{align}
 
\item Ward Identity for \boldmath{$\delta_{L_{-1}}$} and \boldmath{$\delta_{\Bar{L}_{-1}}$} \unboldmath{} 

Now, we will use the results and discussion of section \ref{Leibniz theorem} and demonstrate the simplification thereof. Since we are getting divergent results from terms involving $\partial_{z_{1}}$ and $\partial_{z_{2}}$ we need to re-write the $\partial_{z_{3}}$ term so that we can cancel these divergence. We show this explicitly for all the Lorentz transformations.

We have from $\eqref{L+L-1}$ and $\eqref{L-L-1}$, 
\begin{align}
    \delta_{L_{-1}}\langle L^{-}[\phi_{\Delta_1}]L^{-}[\phi_{\Delta_2}]L^{+}[\phi_{\Delta_{3}}] \rangle=(\partial_{z_{1}}+\partial_{z_{2}}+\partial_{z_{3}})\mathcal{W}(\Delta_{1},\Delta_{2},\Delta_{3}).
\end{align}
Let us explicitly write down the $\partial_{z_{3}}$ term using the definition of light transformation \eqref{L+},
\begin{equation}
    \partial_{z_{3}}\mathcal{W}(\Delta_{1},\Delta_{2},\Delta_{3})=\partial_{z_{3}}\int_{z_{1}}^{z_{2}}dw_{3}\frac{\mathcal{M}(\Delta_{1},\Delta_{2},\Delta_{3})}{(w_{3}-z_{3})^{2-\Delta_{3}}}=\int_{z_{1}}^{z_{2}}dw_{3} \ \mathcal{M}(\Delta_{1},\Delta_{2},\Delta_{3})\partial_{z_{3}}(w_{3}-z_{3})^{\Delta_{3}-2} \label{IntermediateStepL-1}
\end{equation}
where we define $\mathcal{M}(\Delta_{1},\Delta_{2},\Delta_{3})$ as the modified Mellin transformed amplitude,\eqref{3ptScalarMellin}, with $z_{3}$ replaced by $w_{3}$ which follows from the definition of light transformation. Now, we re-write the last term as, 
$$
\partial_{z_{3}}(w_{3}-z_{3})^{\Delta_{3}-2}=-\partial_{w_{3}}(w_{3}-z_{3})^{\Delta_{3}-2}.
$$
Using this in \eqref{IntermediateStepL-1} and performing integration by parts we get,
\begin{equation}
    \partial_{z_{3}}\mathcal{W}(\Delta_{1},\Delta_{2},\Delta_{3})= -\frac{\mathcal{M}(\Delta_{1},\Delta_{2},\Delta_{3})}{(w_{3}-z_{3})^{2-\Delta_{3}}} \Bigg|_{z_{1}}^{z_{2}} + \int_{z_{1}}^{z_{2}}dw_{3}\frac{\partial_{w_{3}}\mathcal{M}(\Delta_{1},\Delta_{2},\Delta_{3})}{(w_{3}-z_{3})^{2-\Delta_{3}}}. \label{DivergenceCancellation}
\end{equation}
The first term exactly cancels the divergent terms arising from \eqref{divergencez1} and \eqref{divergencez2}. Using the three-point function \eqref{3ptScalarMellin} and the above simplification we get the following straightforward expression for the Ward identity, 
\begin{align}
   \delta_{L_{-1}}\langle &L^{-}[\phi_{\Delta_1}]L^{-}[\phi_{\Delta_2}]L^{+}[\phi_{\Delta_{3}}] \rangle=4\pi^{2}\Gamma(\beta-4)(-1)^{\Delta_{1}+\Delta_{2}-2}\Bar{z}^{\Delta_{1}-2}_{31}\Bar{z}^{\Delta_{2}-2}_{32}\nn \\
    & \times \int_{z_{1}}^{z{2}}dw_{3} \Bigg[ z_{12}^{\Delta_{3}-2}(4-\beta)\frac{(z_{2}-w_{3})^{\Delta_{1}-2}(w_{3}-z_{1})^{\Delta_{2}-2}}{(w_{3}-z_{3})^{2-\Delta_{3}}U^{\beta-3}}\Bigg]i(u_{32}+u_{13}+u_{21})=0. \label{3ptL-1Ward}
\end{align}
This proves that the light-transformed three-point function satisfies the Ward identity for $\delta_{L_{-1}}$. Moving forward we check the Ward identity for $\delta_{\Bar{L}_{-1}}$ as well using \eqref{L+Lbar-1} and \eqref{L-Lbar-1}, 
\begin{align}
\delta_{\Bar{L}_{-1}}\langle L^{-}[\phi_{\Delta_1}]L^{-}[\phi_{\Delta_2}]L^{+}[\phi_{\Delta_{3}}] \rangle=\sum_{i=1}^{3}\partial_{\Bar{z}_{i}}\mathcal{W}(\Delta_{1},\Delta_{2},\Delta_{3}).
\end{align}
Since we are not encountering any derivative with respect to $z_{i}$'s  we do not encounter any divergence, so one gets,
\begin{align}
     \delta_{\Bar{L}_{-1}}\langle &L^{-}[\phi_{\Delta_1}]L^{-}[\phi_{\Delta_2}]L^{+}[\phi_{\Delta_{3}}] \rangle=4\pi^{2}\Gamma(\beta-4)(-1)^{\Delta_{1}+\Delta_{2}-2}\Bar{z}^{\Delta_{1}-2}_{31}\Bar{z}^{\Delta_{2}-2}_{32}\int_{z_{1}}^{z{2}}dw_{3} \Bigg[ z_{12}^{\Delta_{3}-2}\nn \\
    & \times \frac{(z_{2}-w_{3})^{\Delta_{1}-2}(w_{3}-z_{1})^{\Delta_{2}-2}}{(w_{3}-z_{3})^{2-\Delta_{3}}U^{\beta-4}}\Bigg]\Big(\Bar{z}_{32}^{-1}(2-\Delta_{2}+\Delta_{2}-2)+\Bar{z}_{31}^{-1}(2-\Delta_{1}+\Delta_{1}-2)\Big)=0. \label{3ptLbar-1Ward}
\end{align}

\end{itemize}

\begin{comment}    
\section{OPE of Light transformed scalar fields}
The operator product expansion plays an important role in conformal field theory since it gives us the algebra of the fields. We start by decomposing the product of indicator functions as in \cite{Banerjee:2022hgc},
$$
\prod_{j=1}^{3}\tau(\sigma^{*}_{j})=\Theta(z_{12})\Theta(z_{13})\Theta(z_{32})+\Theta(z_{21})\Theta(z_{31})\Theta(z_{23}).
$$
Only the second term of the right-hand side contributes to the OPE since we are working with the ordering $z_{1}<z_{3}<z_{2}$
\end{comment}

\subsection{Two point function of Gluons} \label{Section2ptGluon}
The two-point function of gluons has the same form as that for scalars and is given by, 
\begin{align}
    \langle \phi_{z_{1},\Delta_{1}} \phi_{\Bar{z}_{2},\Delta_{2}}\rangle&= 4\pi^{2}\frac{\Gamma(\Delta_{1}+\Delta_{2}-2)}{i^{\Delta_{1}+\Delta_{2}-2}} \frac{\delta^{2}(z_{1}-z_{2})}{(u_{1}-u_{2})^{\Delta_{1}+\Delta_{2}-2}} \label{2ptMellinGluon}
\end{align}
\begin{comment}
    \nn \\
    &=(-1)^{\Delta_{1}+\Delta_{2}}4\pi^{2}\frac{\Gamma(\Delta_{1}+\Delta_{2}-2)}{i^{\Delta_{1}+\Delta_{2}-2}} \frac{\delta^{2}(z_{1}-z_{2})}{(u_{2}-u_{1})^{\Delta_{1}+\Delta_{2}-2}}.
\color{red}We write the second equality to express the fact that the sign of $u_{12}$ is not important for us since all our Ward identities involve \textit{z-derivatives}. \color{black}
\end{comment}
where the index $z$ represents helicity $+1$ and $\Bar{z}$ represents helicity $-1$ and the signs for $\epsilon_{i}$'s are the same as that for \eqref{2ptMellin} also note that we have suppressed the colour indices of the fields. It was shown in \cite{Bagchi:2023cen} that the two-point function computed using all other combination of indices is identically zero either due to the $\delta(z_{12})$ and $\delta(\Bar{z}_{12})$ or due to the fact the $u$ component of the field is a pure gauge term\footnote{ Note that we do not have a factor of $\frac{1}{(1+z_{1}\Bar{z}_{1})(1+z_{2}\Bar{z}_{2})}$ sitting in front of the two-point function. That is because we are following the conventions of \cite{Salzer:2023jqv} where they have shown that the induced metric on $\mathcal{I}^{\pm}$ is flat (take a look at Eq.2.15, Eq. 2.16 and Appendix A). This is consistent with the fact that the Celestial sphere is a Riemann sphere \cite{Stieberger:2018onx} which is conformally equivalent to $\mathbb{R}^{2}$.}.  

The conformal weights of the fields $\phi_{z_{1},\Delta_{1}}(u_{1},z_{1},\Bar{z}_{1})$ are given by $\mathfrak{h}_{1}=(h_{1},\Bar{h}_{1})=(\frac{\Delta_{1}+1}{2},\frac{\Delta_{1}-1}{2})$, similarly for $\phi_{\Bar{z}_{2},\Delta_{2}}(u_{2},z_{2},\Bar{z}_{2})$ we have $\mathfrak{h}_{2}=(h_{2},\Bar{h}_{2})=(\frac{\Delta_{2}-1}{2},\frac{\Delta_{2}+1}{2})$ wherein we used the conventions of \cite{Pasterski:2017ylz}.  The light transformation of \eqref{2ptMellinGluon} using \eqref{L+} and \eqref{L-} gives us, 
\begin{equation}
   \mathcal{G}(\Delta_{1},\Delta_{2})= \langle L^{+}[\phi^{+}_{\Delta_{1}}]L^{-}[\phi^{-}_{\Delta_{2}}]\rangle= 4\pi^{2}\frac{\Gamma(\Delta_{1}+\Delta_{2}-2)}{i^{\Delta_{1}+\Delta_{2}-2}} \frac{(z_{2}-z_{1})^{\Delta_{1}-1}(\Bar{z}_{1}-\Bar{z}_{2})^{\Delta_{2}-1}}{(u_{1}-u_{2})^{\Delta_{1}+\Delta_{2}-2}}. \label{2ptLightGluon}
\end{equation}
where we have replaced $z/\Bar{z}$ by $+/-$ for brevity.

\subsection{Three point function of Gluons} \label{3ptGluon}
The MHV gluon three-point amplitude is 
\begin{align}
    \mathcal{M}(1^{-}2^{-}3^{+})=-8(-1)^{\Delta_{1}+\Delta_{2}}\Gamma(\beta-3)\delta(\Bar{z}_{13})\delta(\Bar{z}_{23})\frac{z_{12}^{\Delta_{3}}z_{23}^{\Delta_{1}-2}z_{31}^{\Delta_{2}-2}}{[i(u_{1}z_{23}+u_{2}z_{31}+u_{3}z_{12})]^{\beta-3}}\Theta\Big(\frac{z_{32}}{z_{12}}\Big)\Theta\Big(\frac{z_{13}}{z_{12}}\Big)  \label{3ptGluonMellin}
\end{align}
where $\beta=\Delta_{1}+\Delta_{2}+\Delta_{3}$. Performing a light transformation on the third field,
\begin{align}
    \langle\phi^{-}_{\Delta_{1}}\phi^{-}_{\Delta_{2}}L^{+}&[\phi^{+}_{\Delta_{3}}]\rangle=-8(-1)^{\Delta_{1}+\Delta_{2}}\ \mathrm{sgn}(z_{21})\Gamma(\beta-3)\delta(\Bar{z}_{13})\delta(\Bar{z}_{23})z_{12}^{\Delta_{3}} \nn \\
    &\times\int_{z_{1}}^{z_{2}}dw_{3}\frac{(z_{2}-w_{3})^{\Delta_{1}-2}(w_{3}-z_{1})^{\Delta_{2}-2}}{(w_{3}-z_{3})^{1-\Delta_{3}}[i(u_{1}(z_{2}-w_{3})+u_{2}(w_{3}-z_{1})+u_{3}z_{12})]^{\beta-3}} \label{3ptGluonThirdLight}
\end{align}
but this does not cure the distributional nature of the correlation functions. We perform light transformations on the first two primaries to get, 
\begin{align}
\langle L^{-}[\phi^{-}_{\Delta_{1}}]L^{-}[\phi^{-}_{\Delta_{2}}]L^{+}[&\phi^{+}_{\Delta_{3}}]\rangle=-8(-1)^{\Delta_{1}+\Delta_{2}}\ \mathrm{sgn}(z_{21})\Gamma(\beta-3)\Bar{z}_{31}^{\Delta_{1}-1}\Bar{z}_{32}^{\Delta_{2}-1}z_{12}^{\Delta_{3}} \nn \\
    &\times \int_{z_{1}}^{z_{2}}dw_{3}\frac{(z_{2}-w_{3})^{\Delta_{1}-2}(w_{3}-z_{1})^{\Delta_{2}-2}(w_{3}-z_{3})^{\Delta_{3}-1}}{[i(u_{1}(z_{2}-w_{3})+u_{2}(w_{3}-z_{1})+u_{3}z_{12})]^{\beta-3}}  \label{3ptMHVGluonAllLight}.
\end{align}
Comparing this with the light-transformed three-point function of the scalars \eqref{3ptScalarLight} one can easily check that they share a similar form, that is, the only difference being in the powers of $z_{ij}$'s. The conformal dimensions of gluon fields are different from those of the scalars. Thus by following the analysis presented in section \ref{2ptScalar} and section \ref{3ptScalar} one can check that the two-point and the three-point correlation function of light-transformed gluon fields satisfy the Ward identities of the Poincar\'e transformations.

Similarly, we give the result of the light-transformed anti-MHV correlation function,
\begin{align}
    \langle L^{+}[\phi^{+}_{\Delta_{1}}]L^{+}[\phi^{+}_{\Delta_{2}}]L^{-}[&\phi^{-}_{\Delta_{3}}]\rangle=-8(-1)^{\Delta_{1}+\Delta_{2}}\ \mathrm{sgn}(\Bar{z}_{21})\Gamma(\beta-3) z_{31}^{\Delta_{1}-1} z_{32}^{\Delta_{2}-1}\Bar{z}_{12}^{\Delta_{3}} \nn \\
    &\times \int_{\Bar{z}_{1}}^{\Bar{z}_{2}}d\Bar{w}_{3}\frac{(\Bar{z}_{2}-\Bar{w}_{3})^{\Delta_{1}-2}(\Bar{w}_{3}-\Bar{z}_{1})^{\Delta_{2}-2}(\Bar{w}_{3}-\Bar{z}_{3})^{\Delta_{3}-1}}{[i(u_{1}(\Bar{z}_{2}-\Bar{w}_{3})+u_{2}(\Bar{w}_{3}-\Bar{z}_{1})+u_{3}\Bar{z}_{12})]^{\beta-3}} \label{3ptantiMHVGluonAllLight}.
\end{align}

As we mentioned at the end of Section \ref{3ptScalar} we made the choice of working with only one channel, namely $12\leftrightharpoons3$. The reason for this choice can be understood from the analysis of Section 4.1 of \cite{Banerjee:2022hgc} where it was shown that the light transformation of the ordinary Mellin $12\leftrightharpoons3$ channel satisfied all the Ward identities of the Poincar\'e transformations. However, it was seen that the other two channels, $13\leftrightharpoons2$ and  $23\leftrightharpoons1$, do not satisfy the Ward identity of Special conformal transformation because such a transformation changes the limits of the integration, involved in performing a light transformation, in a non-trivial manner. This can be understood as effectively taking the three-point function outside its domain of definition using such a transformation. Nevertheless, this symmetry could be recovered by considering a sum over all channels since the factors on the right-hand side of the Ward identity\footnote{Strictly this can not be called a Ward identity since the right-hand side is non-zero, however, we use this nomenclature to avoid verbosity.} for both the channels, $13\leftrightharpoons2$ and  $23\leftrightharpoons1$, were opposite in sign. This is why we choose to work with the $12\leftrightharpoons3$ channel however one could perform a similar investigation for the other two channels keeping in mind the subtlety mentioned above. 

\section{Coincident limit and connecting with Carroll}\label{CarrollConnection}
In this section, we make a direct connection between light-transformed correlation functions and \textit{2D CFT} branch of Carrollian conformal field theories by making a simple observation that the correlation function of \eqref{2ptLight} has the correct scaling behaviour as two-point function in 2D CFT but it does not have the Dirac delta constraint on the conformal weights. 

One could recover the Dirac delta constraint on the weights if one takes a nearly coincident limit, i.e. if one considers all the fields to be inserted on $\mathcal{I^{+}}$ at the same null time. This can be achieved by considering the following limit, $u_{1}\rightarrow u_{2}$. As a consequence one retrieves the 2D CFT two-point function contrary to the Delta function branch, \eqref{2ptMellin}, from which it is not possible to obtain a 2D CFT two-point function by a limiting procedure.
\begin{eqnarray} 
    \boxed{\lim_{u_{1}\rightarrow u_{2}} \langle L^{+}[\phi_{\Delta_{1}}]L^{-}[\phi_{\Delta_{2}}]\rangle =  2\pi^{3} (z_2 -z_1)^{\Delta_1 -2}(\Bar{z}_1-\Bar{z}_2)^{\Delta_{2}-2} \delta(\Delta_{1}+\Delta_{2}-2)}. \label{u1u2lim}
\end{eqnarray}
In writing the above equation we have referred to the representation of the Dirac delta function in terms of the Gamma function as given in \eqref{Diracdelta}, 
\begin{equation*}
    \delta(x)=\lim_{\nu \rightarrow 0}\frac{\Gamma(x)}{\nu^{x}}\quad; x\in \mathbb{C} \quad \text{and} \quad \nu >0
\end{equation*}
however one must be careful while using this representation when $Re(x)=0$ which is the case for the Gamma functon appearing in \eqref{2ptLight}. It turns out that if one takes the limit $\nu \rightarrow 0$, in \eqref{Diracdelta}, carefully then one gets $\frac{1}{2}\delta(Im(x))$ which, for our purpose, turns out to be the Dirac delta function appearing in $\eqref{u1u2lim}$. We refer the reader to Section 4.1 of \cite{Donnay:2020guq} for a detailed analysis of the generalized Dirac delta function. 
The two-point function \eqref{u1u2lim} is the CFT branch of 3-D Carroll CFTs.

Similar results hold for the three-point function of scalars as well, where in the limit $u_{1} \rightarrow u_{2} \rightarrow  u_{3}$ (all the fields inserted at the same null time ) of the light-transformed scalar three-point function,
 \begin{empheq}[box=\fbox]{align}
     \lim_{\substack{u_1 \rightarrow u_2\\ u_2 \rightarrow u_3}}\langle L^{-}[\phi_{\Delta_1}]L^{-}[\phi_{\Delta_2}]L^{+}[\phi_{\Delta_{3}}] \rangle &=-4\pi^{2}\delta(\beta-4)\ \mathrm{sgn}(z_{21})\bar{z}_{31}^{\Delta_{1}-2}\Bar{z}_{32}^{\Delta_{2}-2}\nn \\
    &\times z_{23}^{1-\Delta_{2}}z_{13}^{1-\Delta_{1}}z_{12}^{-1}B(\Delta_{1}-1,\Delta_{2}-1)
 \end{empheq}
 where $B(x,y)=\frac{\Gamma(x)\Gamma(y)}{\Gamma(x+y)}$ is the Euler Beta function and we must mention that, in this equation and in the rest of this section, the result is independent of the order in which we take this limit. One can do a simple power counting to check that this three-point function has the power law behaviour consistent with a 2D CFT three-point function. 

Imposing this limit on gluon correlators \eqref{2ptLightGluon} and \eqref{3ptMHVGluonAllLight} we recover the results of \cite{Sharma:2021gcz, Banerjee:2022hgc},
\begin{equation}
   \boxed{\lim_{u_{1}\rightarrow u_{2}}\langle L^{+}[\phi^{+}_{\Delta_{1}}]L^{-}[\phi^{-}_{\Delta_{2}}]\rangle= 2\pi^{2}\delta(\Delta_{1}+\Delta_{2}-2) (z_{2}-z_{1})^{\Delta_{1}-1}(\Bar{z}_{1}-\Bar{z}_{2})^{\Delta_{2}-1}}
\end{equation}
 and
\begin{empheq}[box=\fbox]{align}
    \lim_{\substack{u_1 \rightarrow u_2\\ u_2 \rightarrow u_3}}\langle L^{-}[\phi^{-}_{\Delta_{1}}]&L^{-}[\phi^{-}_{\Delta_{2}}]L^{+}[\phi^{+}_{\Delta_{3}}]\rangle=\nn \\
    &-4\ \mathrm{sgn}(z_{21})\delta(\beta-3)\Bar{z}_{31}^{\Delta_{1}-1}\Bar{z}_{32}^{\Delta_{2}-1}z_{13}^{1-\Delta_{1}}z_{23}^{\Delta_{1}+\Delta_{3}-2}B(\Delta_{1}-1,\Delta_{2}-1).
\end{empheq}
The fact that the above expression is independent of $u$ is a consequence of energy conservation, this can be understood by taking the nearly coincident limit of the modified Mellin transformation given in \eqref{ModifiedMellin}. 
%that light transformation maps between two seemingly disconnected branches of Carrollian conformal field theories, namely, the electric branch and the magnetic branch. 

Referring to the analysis of this section we propose that light transformation acts as a map between the CFT branch of Carroll field theories and modified Mellin space correlation functions for operators inserted at equal null time. 
It was shown in \cite{Bagchi:2022emh} that the modified Mellin space correlation functions are nothing but the correlation functions of the electric branch of Carrollian conformal field theories. This along with our aforementioned proposal suggests that light transformation maps between the CFT branch and the electric branch of Carrollian conformal field theories for operators inserted at equal null time. This is an important result of this paper.

\section{Operator Product Expansion of Gluons} \label{SectionOPEgluon}
We will focus on the operator product expansion (OPE) of the light-transformed gluon operators in this section (we refer the reader to \cite{Hollands:2023txn} for a nice review of OPEs in the context of quantum field theory). We take the two-pronged approach in the following. We will derive the OPE from symmetry perspectives alone in section \ref{Intrinsic OPE} and then we will reproduce the same singularity by taking the OPE limit of the three-point function in section \ref{colOPE} as a check.

\subsection{Intrinsic approach to OPE of Gluons} \label{Intrinsic OPE}
In this section, we use the non-trivial action of the translation generators to fix the light-transformed OPEs leading pole structure. The working assumption is that the light-transformed theory has the properties of a 2D CFT. Hence, we start with a generic 2D CFT ansatz for an OPE. We assume that the factor of $u_{12}$ does not appear at the leading order in the OPE of the light-transformed theory. This was shown for the OPE of Carrollian fields in \cite{Mason:2023mti} and some of these OPEs were derived by taking an intrinsic approach in \cite{Saha:2023abr}. 

We want to find the singularity structure of the OPE without resorting to bulk physics. We start with the following ansatz for the mixed helicity OPE, 
\begin{align}
    L^{+}[\phi^{+}_{\Delta_{1}}]L^{-}[\phi^{-}_{\Delta_{2}}]=C(\Delta_{1},\Delta_{2})z_{12}^{\Tilde{h}_{p}-\Tilde{h}_{1}-\Tilde{h}_{2}}\Bar{z}_{21}^{\Tilde{\Bar{h}}_{p}-\Tilde{\Bar{h}}_{1}-\Tilde{\Bar{h}}_{2}}L^{-}[\phi^{-}_{\Delta_{p}}] 
\end{align}
where $\Tilde{h}$ and $\Tilde{\Bar{h}}$ are the weights of the light-transformed fields defined in \eqref{L+} and \eqref{L-}. It should be noted that since $\Delta_p$ is arbitrary, this is the most general operator that could contribute to the OPE at the leading order\footnote{Another possibility is contribution from $L^{+}[\phi^{+}_{\Delta_{p}}]$ which we consider in \eqref{IntrinsicTotalgluonOPE2}.}. We have not explicitly written the sum over the descendants of the operator since we are interested in the leading term only. We note that $\epsilon_{1}=-\epsilon_{2}=-\epsilon_{p}=-1$ and re-write the above equation in terms of the scaling $\Delta_{i}$ of the fields.
\begin{align}
    L^{+}[\phi^{+}_{\Delta_{1}}]L^{-}[\phi^{-}_{\Delta_{2}}]=C(\Delta_{1},\Delta_{2})z_{12}^{\frac{\Delta_{p}+\Delta_{1}-\Delta_{2}-1}{2}}\Bar{z}_{21}^{\frac{1+\Delta_{2}-\Delta_{p}-\Delta_{1}}{2}}L^{-}[\phi^{-}_{\Delta_{p}}] .\label{OPEansatz}
\end{align}
One knows from the analysis of 2D CFT that we have already exhausted the global conformal symmetries to fix the above mentioned form of the OPE and that no further information can be extracted from those generators. Hence, we will fix $\Delta_{p}$ using the translation symmetry which amounts to fixing the singularity structure of the OPE. We start with the action of $\delta_{M_{00}}$ given in \eqref{L+M00} and \eqref{L-M00} and get, 
$$
    \delta_{M_{00}}L^{+}[\phi^{+}_{\Delta_{1}}]L^{-}[\phi^{-}_{\Delta_{2}}]+L^{+}[\phi^{+}_{\Delta_{1}}]\delta_{M_{00}}L^{-}[\phi^{-}_{\Delta_{2}}]=C(\Delta_{1},\Delta_{2})z_{12}^{\frac{\Delta_{p}+\Delta_{1}-\Delta_{2}-1}{2}}\Bar{z}_{21}^{\frac{1+\Delta_{2}-\Delta_{p}-\Delta_{1}}{2}}\delta_{M_{00}}L^{-}[\phi^{-}_{\Delta_{p}}]
$$
Using the OPE ansatz and collecting similar powers of $z_{12}$ and $\Bar{z}_{21}$ we get,
\begin{align}
    &\Bigg[-\frac{C(\Delta_{1}+1,\Delta_{2})}{\Delta_{1}}\bigg( \frac{\Delta_{p}+\Delta_{1}-\Delta_{2}+1}{2}\bigg)  + \frac{C(\Delta_{1},\Delta_{2}+1)}{\Delta_{2}}\bigg( \frac{1+\Delta_{2}-\Delta_{p}-\Delta_{1}}{2}\bigg)\Bigg] \nn \\
    & \times z_{12}^{\frac{\Delta_{p}+\Delta_{1}-\Delta_{2}-1}{2}}\Bar{z}_{21}^{\frac{\Delta_{2}-\Delta_{p}-\Delta_{1}-1}{2}}L^{-}[\phi^{-}_{\Delta_{p}+1}] +\frac{C(\Delta_{1},\Delta_{2}+1)}{\Delta_{2}}z_{12}^{\frac{\Delta_{p}+\Delta_{1}-\Delta_{2}-1}{2}}\Bar{z}_{21}^{\frac{1+\Delta_{2}-\Delta_{p}-\Delta_{1}}{2}}\partial_{\Bar{z}_{2}}L^{-}[\phi^{-}_{\Delta_{p}+1}] \nn \\
    &=\frac{C(\Delta_{1},\Delta_{2})}{\Delta_{p}}z_{12}^{\frac{\Delta_{p}+\Delta_{1}-\Delta_{2}-1}{2}}\Bar{z}_{21}^{\frac{1+\Delta_{2}-\Delta_{p}-\Delta_{1}}{2}}\partial_{\Bar{z}_{2}}L^{-}[\phi^{-}_{\Delta_{p}+1}].
\end{align}
Comparing the powers of $z_{12}$ and $\Bar{z}_{21}$ we get the following equations for $C(\Delta_{1},\Delta_{2})$ and $\Delta_{p}$,
\begin{align}
    &\frac{C(\Delta_{1}+1,\Delta_{2})}{\Delta_{1}}\bigg( \frac{\Delta_{p}+\Delta_{1}-\Delta_{2}+1}{2}\bigg)  -\frac{C(\Delta_{1},\Delta_{2}+1)}{\Delta_{2}}\bigg( \frac{1+\Delta_{2}-\Delta_{p}-\Delta_{1}}{2}\bigg)=0 \label{Recursion1} \\
    &\frac{C(\Delta_{1},\Delta_{2}+1)}{\Delta_{2}}=\frac{C(\Delta_{1},\Delta_{2})}{\Delta_{p}}. \label{Recursion2}
\end{align}
Similarly, using the $\delta_{M_{01}}$ symmetry generator given in \eqref{L-M01} and \eqref{L+M01} we get, 
\begin{align}
    &\Bigg[-\frac{\Bar{z}_{1}C(\Delta_{1}+1,\Delta_{2})}{\Delta_{1}}\bigg(\frac{\Delta_{p}+\Delta_{1}-\Delta_{2}+1}{2}\bigg)+ \frac{\Bar{z}_{2}C(\Delta_{1},\Delta_{2}+1)}{\Delta_{2}}\bigg( \frac{1+\Delta_{2}-\Delta_{p}-\Delta_{1}}{2}\bigg)\Bigg] \Bigg] \nn \\
    & \times z_{12}^{\frac{\Delta_{p}+\Delta_{1}-\Delta_{2}-1}{2}}\Bar{z}_{21}^{\frac{\Delta_{2}-\Delta_{p}-\Delta_{1}-1}{2}}L^{-}[\phi^{-}_{\Delta_{p}+1}] + \frac{C(\Delta_{1},\Delta_{2}+1)}{\Delta_{2}}z_{12}^{\frac{\Delta_{p}+\Delta_{1}-\Delta_{2}-1}{2}}\Bar{z}_{21}^{\frac{1+\Delta_{2}-\Delta_{p}-\Delta_{1}}{2}}\Bar{z}_{2}\partial_{\Bar{z}_{2}}L^{-}[\phi^{-}_{\Delta_{p}+1}] \nn \\
    & -C(\Delta_{1},\Delta_{2}+1)z_{12}^{\frac{\Delta_{p}+\Delta_{1}-\Delta_{2}-1}{2}}\Bar{z}_{21}^{\frac{1+\Delta_{2}-\Delta_{p}-\Delta_{1}}{2}}L^{-}[\phi^{-}_{\Delta_{p}+1}]\nn \\ \nn\\
    & =-\frac{C(\Delta_{1},\Delta_{2})}{\Delta_{p}}z_{12}^{\frac{\Delta_{p}+\Delta_{1}-\Delta_{2}-1}{2}}\Bar{z}_{21}^{\frac{1+\Delta_{2}-\Delta_{p}-\Delta_{1}}{2}}\bigg(-\frac{\Bar{z}_{2}}{\Delta_{p}}\partial_{\Bar{z}_{2}} +1\bigg)L^{-}[\phi^{-}_{\Delta_{p}+1}].
\end{align}
Rewriting $\Bar{z}_{1}=\Bar{z}_{12}+\Bar{z}_{2}$ we get the following equation along with \eqref{Recursion1} and \eqref{Recursion2}, 
\begin{align}
    C(\Delta_{1},\Delta_{2}+1)-\frac{C(\Delta_{1}+1,\Delta_{2})}{\Delta_{1}}\bigg(\frac{\Delta_{p}+\Delta_{1}-\Delta_{2}+1}{2}\bigg)=C(\Delta_{1},\Delta_{2}). \label{recursion3}
\end{align}
Now we will simultaneously solve \eqref{Recursion1},\eqref{Recursion2} and \eqref{recursion3}. Eliminating $\bigg(\frac{\Delta_{p}+\Delta_{1}-\Delta_{2}+1}{2} \frac{C(\Delta_{1}+1,\Delta_{2})}{\Delta_{1}}\bigg)$ from \eqref{Recursion1} using \eqref{recursion3} we get, 
$$
C(\Delta_{1},\Delta_{2}+1)-C(\Delta_{1},\Delta_{2})-\frac{C(\Delta_{1},\Delta_{2}+1)}{\Delta_{2}}\bigg( \frac{1+\Delta_{2}-\Delta_{p}-\Delta_{1}}{2}\bigg)=0
$$
finally eliminating $C(\Delta_{1},\Delta_{2})$ from the above equation using \eqref{Recursion2} we get, 
\begin{equation}
    C(\Delta_{1},\Delta_{2}+1)\bigg[ 1-\frac{\Delta_{p}}{\Delta_{2}}-\frac{1+\Delta_{2}-\Delta_{p}-\Delta_{1}}{2\Delta_{2}}\bigg]=0.
\end{equation}
Lastly, demanding that $C(\Delta_{1},\Delta_{2}+1)\neq 0$ we get, 
\begin{equation}
    \Delta_{p}=\Delta_{1}+\Delta_{2}-1. \label{OPEweight}
\end{equation}
One can see that using the expression of $\Delta_{p}$ in our ansatz for the leading term \eqref{OPEansatz} one gets the following singularity structure, 
\begin{equation}
    \boxed{L^{+}[\phi^{+}_{\Delta_{1}}]L^{-}[\phi^{-}_{\Delta_{2}}]=C(\Delta_{1},\Delta_{2})z_{12}^{\Delta_{1}-1}\Bar{z}_{21}^{1-\Delta_{1}}L^{-}[\phi^{-}_{\Delta_{p}}]}.
\end{equation}
Thus we can see that one can fix the leading singularity of the OPE using the symmetries of the theory when working in the light operator basis. We can solve for $C(\Delta_{1},\Delta_{2})$ using $\Delta_{p}$ in \eqref{Recursion1}, \eqref{Recursion2} and \eqref{recursion3}. We get the following recursion relations for $C(\Delta_{1},\Delta_{2})$,
\begin{align}
 &C(\Delta_{1},\Delta_{2}+1)-C(\Delta_{1}+1,\Delta_{2})=C(\Delta_{1},\Delta_{2})   \\
 &(\Delta_{1}+\Delta_{2}-1)C(\Delta_{1},\Delta_{2}+1)=\Delta_{2}C(\Delta_{1},\Delta_{2})\\
 &(1-\Delta_{1}-\Delta_{2})C(\Delta_{1}+1,\Delta_{2})=(\Delta_{1}-1)C(\Delta_{1},\Delta_{2}) 
\end{align}

It is known from the analysis of \cite{Pate:2019mfs} that the OPE coefficient of a mixed helicity gluon OPE in Celestial CFT do not have symmetry under exchange of $\Delta_{1}$ and $\Delta_{2}$, that is, $C(\Delta_{1},\Delta_{2})\neq C(\Delta_{2},\Delta_{1})$. Using Weiland's theorem \cite{db6dc5fa-41f3-3c6a-acf0-210f0aa62d5a} we can solve for $C(\Delta_{1},\Delta_{2})$,
\begin{equation}
    C(\Delta_{1},\Delta_{2})=-C(2,1)\frac{\Gamma(\Delta_{1}-1)\Gamma(2-\Delta_{1}-\Delta_{2})}{\Gamma(1-\Delta_{2})}=-C(2,1)B(\Delta_{1}-1,2-\Delta_{1}-\Delta_{2})
\end{equation}
where $C(2,1)$ is a normalization constant. We will show that this constant can be fixed using the soft limit of the three-point correlator. In terms of the unfixed constant we have, 
\begin{equation}
 L^{+}[\phi^{+}_{\Delta_{1}}]L^{-}[\phi^{-}_{\Delta_{2}}]=-C(2,1)B(\Delta_{1}-1,2-\Delta_{1}-\Delta_{2})z_{12}^{\Delta_{1}-1}\Bar{z}_{21}^{1-\Delta_{1}}L^{-}[\phi^{-}_{\Delta_{1}+\Delta_{2}-1}]. \label{UnfixedOPE1}  
\end{equation}
 In the next section, we will fix the normalization constant by studying the soft limit of the light-transformed three-point function of gluons given in equations \eqref{3ptMHVGluonAllLight} and \eqref{3ptantiMHVGluonAllLight}.
\subsection{Soft Limit of the three-point function} \label{SoftGluon}
In this subsection, we study the soft limit of the gluon three-point function given in \eqref{3ptGluonMellin}, using which we will derive the soft limit of the light-transformed gluon correlator. We will use the results of this section to fix the normalization constant appearing while calculating the OPE coefficient using the symmetries of our theory \cite{Pate:2019lpp}. 
This channel i.e. $12\leftrightharpoons 3$ is non-zero for $z_{21}>0$ or $z_{21}<0$; we choose to work with $z_{21}>0$ (or $(\Bar{z}_{21})>0$ for the case of anti-MHV amplitudes) henceforth and then state the result for $z_{21}<0$ (or $(\Bar{z}_{21})<0$ for the case of anti-MHV amplitudes) since the analysis is similar for both the orderings. 

%The origin of the sign can be understood by noticing that the light-transformed three-point function in Celestial CFT is itself defined up to a sign. Since the leading term in the OPE is expected to contribute to the three-point function, the fixing of OPE coefficients up to a sign is reasonable. \color{black} 

We begin by noting down an important representation of the Dirac delta function as 
\begin{equation}
    \delta(x)=\lim_{\epsilon\rightarrow0}\frac{\epsilon}{2}|x|^{\epsilon-1}.
\end{equation}
We will use this to simplify our calculations. We take the second particle to be the soft one and get, 
\begin{align}
    \lim_{(\Delta_{2}-1)\rightarrow 0}(\Delta_{2}-1)  &\langle\phi^{-}_{\Delta_{1}}\phi^{-}_{\Delta_{2}}\phi^{+}_{\Delta_{3}}\rangle=\lim_{(\Delta_{2}-1)\rightarrow 0}(\Delta_{2}-1)z_{31}^{\Delta_{2}-2}  \nn \\
&\times \Bigg(\frac{-8(-1)^{\Delta_{1}+\Delta_{2}}\Gamma(\Delta_{1}+\Delta_{2}+\Delta_{3}-3)\delta(\Bar{z}_{13})\delta(\Bar{z}_{23})z_{12}^{\Delta_{3}}z_{23}^{\Delta_{1}-2}}{[i(u_{1}z_{23}+u_{2}z_{31}+u_{3}z_{12})]^{\Delta_{1}+\Delta_{2}+\Delta_{3}-3}}\Theta\Big(\frac{z_{32}}{z_{12}}\Big)\Theta\Big(\frac{z_{13}}{z_{12}}\Big)\Bigg).
\end{align}
The term outside the big brackets gives us $2\ \delta(z_{31})$ and using this one can simplify the expression inside the bracket to get, 
\begin{align}
    \lim_{(\Delta_{2}-1)\rightarrow 0}(\Delta_{2}-1)\langle\phi^{-}_{\Delta_{1}}\phi^{-}_{\Delta_{2}}&\phi^{+}_{\Delta_{3}}\rangle=\frac{8(-1)^{\Delta_{1}+\Delta_{3}}\Gamma(\Delta_{1}+\Delta_{3}-2)}{(iu_{13})^{\Delta_{1}+\Delta_{3}-2}}\delta({z_{13}})\delta(\Bar{z}_{13})\delta(\Bar{z}_{23}). 
\end{align}
Re-writing this in terms of the two-point function one gets, 
\begin{align}
   \lim_{(\Delta_{2}-1)\rightarrow 0}(\Delta_{2}-1)\langle\phi^{-}_{\Delta_{1}}\phi^{-}_{\Delta_{2}}&\phi^{+}_{\Delta_{3}}\rangle=\frac{2}{\pi^{2}}\delta(\Bar{z}_{23})\langle \phi_{\Delta_{1}}^{-}\phi_{\Delta_{3}}^{+}\rangle   \label{SoftGluonMHVMellin} 
\end{align}
wherein we used the factor of $(-1)^{\Delta_{1}+\Delta_{3}}$ in writing the two-point function \eqref{2ptMellinGluon}. One can immediately see that this kind of soft factorization is in contrast with the soft factorization obtained in \cite{Pate:2019mfs}. That is because the soft factorization studied in \cite{Pate:2019mfs} is for correlation functions with four or more field insertions. Now, following equation \eqref{3ptMHVGluonAllLight} and performing three light transformations on both sides we get,
\begin{equation}
    \lim_{(\Delta_{2}-1)\rightarrow 0}(\Delta_{2}-1)\langle L^{-}[\phi^{-}_{\Delta_{1}}]L^{-}[\phi^{-}_{\Delta_{2}}]L^{+}[\phi^{+}_{\Delta_{3}}]\rangle=\frac{2}{\pi^{2}}\langle L^{-}[\phi_{\Delta_{1}}^{-}]L^{+}[\phi_{\Delta_{3}}^{+}]\rangle. \label{SoftGluonMHVLight}
\end{equation}
This result can be obtained by taking the conformally soft limit in \eqref{3ptMHVGluonAllLight} as well, for that one needs to rewrite equation \eqref{3ptantiMHVGluonAllLight} in terms of the step functions as follows,
\begin{align}
    \langle L^{-}[\phi^{-}_{\Delta_{1}}]L^{-}[&\phi^{-}_{\Delta_{2}}]L^{+}[\phi^{+}_{\Delta_{3}}]\rangle=-8\Gamma(\beta-3)\Bar{z}_{31}^{\Delta_{1}-1}\Bar{z}_{32}^{\Delta_{2}-1}z_{12}^{\Delta_{3}} \nn \\
    &\times \int  dw_{3}\frac{(w_{3}-z_{2})^{\Delta_{1}-2}(z_{1}-w_{3})^{\Delta_{2}-2}(w_{3}-z_{3})^{\Delta_{3}-1}}{[i(u_{1}(z_{2}-w_{3})+u_{2}(w_{3}-z_{1})+u_{3}z_{12})]^{\beta-3}}\Theta\Big(\frac{w_{3}-z_{2}}{z_{12}}\Big)\Theta\Big(\frac{z_{1}-w_{3}}{z_{12}}\Big).
\end{align}
Taking the conformally soft limit one obtains the Dirac delta function $2\delta(w_{3}-z_{1})$. Using this and simplifying the above expression one gets back \eqref{SoftGluonMHVLight}.

Similarly, one can show that if we start from the anti-MHV amplitude and take the soft limit, then we get the following soft factorization
\begin{align}
     \lim_{(\Delta_{2}-1)\rightarrow 0}(\Delta_{2}-1)\langle\phi^{+}_{\Delta_{1}}\phi^{+}_{\Delta_{2}}&\phi^{-}_{\Delta_{3}}\rangle=\frac{2}{\pi^{2}}\delta(z_{23})\langle \phi_{\Delta_{1}}^{+}\phi_{\Delta_{3}}^{-}\rangle \label{SoftGluonantiMHVMellim}
\end{align}
for the Mellin operators and
\begin{equation}
    \lim_{(\Delta_{2}-1)\rightarrow 0}(\Delta_{2}-1)\langle L^{+}[\phi^{+}_{\Delta_{1}}]L^{+}[\phi^{+}_{\Delta_{2}}]L^{-}[\phi^{-}_{\Delta_{3}}]\rangle=\frac{2}{\pi^{2}}\langle L^{+}[\phi_{\Delta_{1}}^{+}]L^{-}[\phi_{\Delta_{3}}^{-}]\rangle. \label{SoftGluonantiMHVLight}
\end{equation}
An important question at this point is, do light-transformed higher point functions admit a soft factorization? We try to address this question in Appendix \ref{SoftLightFour}.

Using equation \eqref{SoftGluonMHVLight} we have, 
\begin{align}
    \lim_{\Delta_{1}-1 \rightarrow 0}(\Delta_{1}-1)L^{+}[\phi^{+}_{\Delta_{1}}]L^{-}[\phi^{-}_{\Delta_{2}}]=\frac{2}{\pi^{2}}\frac{1}{\Delta_{1}-1}L^{+}[\phi^{+}_{\Delta_{2}}] \label{OPEsoft}.
\end{align}
wherein one must keep in mind that $z_{1}>z_{2}$. Taking the soft limit in \eqref{UnfixedOPE1} and comparing with \eqref{OPEsoft} one can conclude that, 
\begin{equation}
    -C(2,1)=\frac{2}{\pi^{2}}
\end{equation}
So, the OPE for the ordering $z_{1}>z_{2}$ is, 
\begin{align}
    \boxed{L^{+}[\phi^{+}_{\Delta_{1}}]L^{-}[\phi^{-}_{\Delta_{2}}]=\frac{2}{\pi^{2}}B(\Delta_{1}-1,2-\Delta_{1}-\Delta_{2})z_{12}^{\Delta_{1}-1}\Bar{z}_{21}^{1-\Delta_{1}}L^{-}[\phi^{-}_{\Delta_{1}+\Delta_{2}-1}]}. \label{intrinsicgluonOPEMHV}
\end{align}
This is one of the main results of this paper. One can similarly obtain the OPE for $L^{-}[\phi^{-}_{\Delta_{1}}]L^{+}[\phi^{+}_{\Delta_{2}}]$ as,
\begin{equation}
    \boxed{L^{-}[\phi^{-}_{\Delta_{1}}]L^{+}[\phi^{+}_{\Delta_{2}}]=\frac{2}{\pi^{2}}B(\Delta_{1}-1,2-\Delta_{1}-\Delta_{2})z_{12}^{1-\Delta_{1}}\Bar{z}_{21}^{\Delta_{1}-1}L^{+}[\phi^{+}_{\Delta_{1}+\Delta_{2}-1}]}. \label{intrinsicgluonOPEantiMHV}
\end{equation}
Interchanging $\Delta_{1}$ and $\Delta_{2}$ in the above equation and summing the right-hand side gives us the full OPE for mixed helicity light-transformed operators, for the ordering $z_{1}>z_{2}$, 
\begin{align}
 L^{+}[\phi^{+}_{\Delta_{1}}]L^{-}[\phi^{-}_{\Delta_{2}}]&=\frac{2}{\pi^{2}}B(\Delta_{1}-1,2-\Delta_{1}-\Delta_{2})z_{12}^{\Delta_{1}-1}\Bar{z}_{21}^{1-\Delta_{1}}L^{-}[\phi^{-}_{\Delta_{1}+\Delta_{2}-1}] \nn \\
    &+\frac{2}{\pi^{2}}B(\Delta_{2}-1,2-\Delta_{1}-\Delta_{2})z_{12}^{1-\Delta_{2}}\Bar{z}_{21}^{\Delta_{2}-1}L^{+}[\phi^{+}_{\Delta_{1}+\Delta_{2}-1}]. \label{IntrinsicTotalgluonOPE}
\end{align}
For readers familiar with the details, we note that, as expected the soft poles corresponding to both the fields are present in the OPE due to the beta function. Similarly, for the other ordering, $z_{1}<z_{2}$, we get, 
\begin{align}
 L^{+}[\phi^{+}_{\Delta_{1}}]L^{-}[\phi^{-}_{\Delta_{2}}]&=-\frac{2}{\pi^{2}}B(\Delta_{1}-1,2-\Delta_{1}-\Delta_{2})z_{12}^{\Delta_{1}-1}\Bar{z}_{21}^{1-\Delta_{1}}L^{-}[\phi^{-}_{\Delta_{1}+\Delta_{2}-1}] \nn \\
    &-\frac{2}{\pi^{2}}B(\Delta_{2}-1,2-\Delta_{1}-\Delta_{2})z_{12}^{1-\Delta_{2}}\Bar{z}_{21}^{\Delta_{2}-1}L^{+}[\phi^{+}_{\Delta_{1}+\Delta_{2}-1}]. \label{IntrinsicTotalgluonOPE2}
\end{align}
In the next section, we will show that the OPE we obtained intrinsically from symmetry arguments matches the OPE obtained by taking the collinear limit of the light-transformed three-point function.

\subsection{OPE from the collinear limit of light-transformed correlation function}\label{colOPE}
In this section, we want to calculate the OPE of two light-transformed operators starting from the light-transformed three-point function which is given by, 
\begin{align}
    \langle L^{-}[\phi^{-}_{\Delta_{1}}(z_{1}\Bar{z}_{1})]L^{-}[\phi^{-}_{\Delta_{2}}(z_{2}\Bar{z}_{2})]&L^{+}[\phi^{+}_{\Delta_{3}}(z_{3}\Bar{z}_{3})]\rangle=-8(-1)^{\Delta_{1}+\Delta_{2}}\ \Gamma(\beta-3)\Bar{z}_{31}^{\Delta_{1}-1}\Bar{z}_{32}^{\Delta_{2}-1}z_{12}^{\Delta_{3}} \nn \\
    &\int_{z_{1}}^{z_{2}}dw_{3}\frac{(z_{2}-w_{3})^{\Delta_{1}-2}(w_{3}-z_{1})^{\Delta_{2}-2}(w_{3}-z_{3})^{\Delta_{3}-1}}{[i(u_{1}(z_{2}-w_{3})+u_{2}(w_{3}-z_{1})+u_{3}z_{12})]^{\beta-3}}.
\end{align}
Note that there is no sign function because of the choice of ordering. The OPE limit corresponds to expanding the above correlation function about $u_{23}=0$, $\Bar{z}_{23}=0$ and $z_{23}=0$ keeping the leading order terms. We simplify the integral by first expanding the denominator about $u_{23}=0$ and get, 
\begin{align}
    \langle L^{-}[\phi^{-}_{\Delta_{1}}(z_{1}\Bar{z}_{1})]L^{-}[\phi^{-}_{\Delta_{2}}(z_{2}\Bar{z}_{2})]&L^{+}[\phi^{+}_{\Delta_{3}}(z_{3}\Bar{z}_{3})]\rangle=-8(-1)^{\Delta_{1}+\Delta_{2}}\ \Gamma(\beta-3)\Bar{z}_{31}^{\Delta_{1}-1}\Bar{z}_{32}^{\Delta_{2}-1}z_{12}^{\Delta_{3}} \nn \\
    &\int_{z_{1}}^{z_{2}}dw_{3}\frac{(z_{2}-w_{3})^{\Delta_{1}-2}(w_{3}-z_{1})^{\Delta_{2}-2}}{(w_{3}-z_{3})^{1-\Delta_{3}}[i(u_{1}(z_{2}-w_{3})+u_{3}(w_{3}-z_{2})]^{\beta-3}}
\end{align}
The denominator can be further simplified by taking $(z_{2}-w_{3})$ outside the brackets and we get, 
\begin{align}
    \langle L^{-}[\phi^{-}_{\Delta_{1}}(z_{1}\Bar{z}_{1})]L^{-}[\phi^{-}_{\Delta_{2}}(z_{2}\Bar{z}_{2})]L^{+}[\phi^{+}_{\Delta_{3}}(z_{3}\Bar{z}_{3})]&\rangle=-8(-1)^{\Delta_{1}+\Delta_{2}}\ \Gamma(\beta-3)\Bar{z}_{31}^{\Delta_{1}-1}\Bar{z}_{32}^{\Delta_{2}-1}z_{12}^{\Delta_{3}} \nn \\
    &\int_{z_{1}}^{z_{2}}dw_{3}\frac{(z_{2}-w_{3})^{-\Delta_{2}-\Delta_{3}+1}(w_{3}-z_{1})^{\Delta_{2}-2}}{(w_{3}-z_{3})^{1-\Delta_{3}}(iu_{13})^{\beta-3}}
\end{align}
Performing the integral gives us, 
$$
\int_{z_{1}}^{z_{2}}dw_{3}\frac{(z_{2}-w_{3})^{-\Delta_{2}-\Delta_{3}+1}(w_{3}-z_{1})^{\Delta_{2}-2}}{(w_{3}-z_{3})^{1-\Delta_{3}}}=\frac{\Gamma(\Delta_{2}-1)\Gamma(2-\Delta_{2}-\Delta_{3})}{\Gamma(1-\Delta_{3})}z_{21}^{-\Delta_{3}}z_{13}^{\Delta_{2}+\Delta_{3}-2}z_{23}^{1-\Delta_{2}}
$$

Plugging this back into the three-point function we get, 
\begin{align}
    \langle L^{-}[\phi^{-}_{\Delta_{1}}(z_{1},\Bar{z}_{1})]&L^{-}[\phi^{-}_{\Delta_{2}}(z_{2},\Bar{z}_{2})]L^{+}[\phi^{+}_{\Delta_{3}}(z_{3},\Bar{z}_{3})]\rangle=-8(-1)^{\Delta_{1}+\Delta_{2}}\ \Gamma(\beta-3)\Bar{z}_{31}^{\Delta_{1}-1}\Bar{z}_{32}^{\Delta_{2}-1}z_{12}^{\Delta_{3}} \nn \\
    &\times \frac{\Gamma(\Delta_{2}-1)\Gamma(2-\Delta_{2}-\Delta_{3})(iu_{13})^{\Delta_{1}+(\Delta_{2}+\Delta_{3}-1)-2}}{\Gamma(1-\Delta_{3})}z_{21}^{-\Delta_{3}}z_{13}^{\Delta_{2}+\Delta_{3}-2}z_{23}^{1-\Delta_{2}}
\end{align}
where the power of $u_{13}$ is suggestively written to compare it with the two-point function. Simplifying this expression one gets, 
\begin{align}
    \langle L^{-}[\phi^{-}_{\Delta_{1}}(z_{1},\Bar{z}_{1})]&L^{-}[\phi^{-}_{\Delta_{2}},(z_{2}\Bar{z}_{2})]L^{+}[\phi^{+}_{\Delta_{3}}(z_{3},\Bar{z}_{3})]\rangle=\frac{2}{\pi^{2}}\frac{\Gamma(\Delta_{2}-1)\Gamma(2-\Delta_{2}-\Delta_{3})}{\Gamma(1-\Delta_{3})}\frac{\Bar{z}_{32}^{\Delta_{2}-1}}{z_{23}^{\Delta_{2}-1}}\nn \\
    &\times \Bigg(4\pi^{2}(-1)^{\Delta_{1}+\Delta_{2}+\Delta_{3}-1}\Gamma(\Delta_{1}+\Delta_{2}+\Delta_{3}-1-2)\frac{\Bar{z}_{31}^{\Delta_{1}-1}z_{13}^{(\Delta_{2}+\Delta_{3}-1)-1}}{(iu_{13})^{\Delta_{1}+(\Delta_{2}+\Delta_{3}-1)-2}}\Bigg)
\end{align}
wherein we have re-arranged the powers of $(-1)$ to write the two-point function. The term in the brackets is nothing but $\langle L^{-}[\phi^{-}_{\Delta_{1}}]L^{+}[\phi^{+}_{\Delta_{2}+\Delta_{3}-1}] \rangle$ , so that we have, 
\begin{align}
    \langle L^{-}[\phi^{-}_{\Delta_{1}}(z_{1}\Bar{z}_{1})]L^{-}[\phi^{-}_{\Delta_{2}}(z_{2}\Bar{z}_{2})]L^{+}[\phi^{+}_{\Delta_{3}}(z_{3}\Bar{z}_{3})]\rangle=\ &\frac{2}{\pi^{2}}\frac{\Gamma(\Delta_{2}-1)\Gamma(2-\Delta_{2}-\Delta_{3})}{\Gamma(1-\Delta_{3})}\frac{\Bar{z}_{32}^{\Delta_{2}-1}}{z_{23}^{\Delta_{2}-1}}\nn \\
    &\times \langle L^{-}[\phi^{-}_{\Delta_{1}}]L^{+}[\phi^{+}_{\Delta_{2}+\Delta_{3}-1}] \rangle
\end{align}

The OPE can be read off from this equation as, 
\begin{equation}
    L^{-}[\phi^{-}_{\Delta_{2}}]L^{+}[\phi^{+}_{\Delta_{3}}]=\frac{2}{\pi^{2}}\frac{\Gamma(\Delta_{2}-1)\Gamma(2-\Delta_{2}-\Delta_{3})}{\Gamma(1-\Delta_{3})}\frac{\Bar{z}_{32}^{\Delta_{2}-1}}{z_{23}^{\Delta_{2}-1}}L^{+}[\phi^{+}_{\Delta_{2}+\Delta_{3}-1}] \label{gluonOPEMHV}
\end{equation}
If one performs a similar analysis for an anti-MHV three-point correlator of gluons keeping $\epsilon_{1}=\epsilon_{2}=-\epsilon_{3}=-1$ \eqref{3ptantiMHVGluonAllLight}, then one gets the following result,
\begin{align}
    \langle L^{+}[\phi^{+}_{\Delta_{1}}(z_{1}\Bar{z}_{1})]L^{+}[\phi^{+}_{\Delta_{2}}(z_{2}\Bar{z}_{2})]&L^{-}[\phi^{-}_{\Delta_{3}}(z_{3}\Bar{z}_{3})]\rangle=-8(-1)^{\Delta_{1}+\Delta_{2}}\ \mathrm{sgn}(\Bar{z}_{21})\Gamma(\beta-3) z_{31}^{\Delta_{1}-1}z_{32}^{\Delta_{2}-1}\Bar{z}_{12}^{\Delta_{3}} \nn \\
    &\int_{\Bar{z}_{1}}^{\Bar{z}_{2}}d\Bar{w}_{3}\frac{(\Bar{z}_{2}-\Bar{w}_{3})^{\Delta_{1}-2}(\Bar{w}_{3}-\Bar{z}_{1})^{\Delta_{2}-2}(\Bar{w}_{3}-\Bar{z}_{3})^{1-\Delta_{3}}}{[i(u_{1}(\Bar{z}_{2}-\Bar{w}_{3})+u_{2}(\Bar{w}_{3}-\Bar{z}_{1})+u_{3}\Bar{z}_{12})]^{\beta-3}}.
\end{align}
Taking the OPE limit we get, 
\begin{equation}
    L^{+}[\phi^{+}_{\Delta_{2}}]L^{-}[\phi^{-}_{\Delta_{3}}]=\frac{2}{\pi^{2}}\frac{\Gamma(\Delta_{2}-1)\Gamma(2-\Delta_{2}-\Delta_{3})}{\Gamma(1-\Delta_{3})}\frac{z_{23}^{\Delta_{2}-1}}{\Bar{z}_{32}^{\Delta_{2}-1}}L^{-}[\phi^{-}_{\Delta_{2}+\Delta_{3}-1}]. \label{gluonOPEantiMHV}
\end{equation}
One can see that \eqref{gluonOPEMHV} and \eqref{gluonOPEantiMHV} are exactly similar to \eqref{intrinsicgluonOPEantiMHV} and \eqref{intrinsicgluonOPEMHV}. Again, interchanging $\Delta_{2}$ and $\Delta_{3}$ in \eqref{gluonOPEantiMHV} and summing with \eqref{gluonOPEMHV} one gets the total OPE as,
\begin{align}
     L^{-}[\phi^{-}_{\Delta_{2}}]L^{+}[\phi^{+}_{\Delta_{3}}]&=\frac{2}{\pi^{2}}\frac{\Gamma(\Delta_{2}-1)\Gamma(2-\Delta_{2}-\Delta_{3})}{\Gamma(1-\Delta_{3})}\frac{\Bar{z}_{32}^{\Delta_{2}-1}}{z_{23}^{\Delta_{2}-1}}L^{+}[\phi^{+}_{\Delta_{2}+\Delta_{3}-1}]\nn \\
     &+ \frac{2}{\pi^{2}}\frac{\Gamma(\Delta_{3}-1)\Gamma(2-\Delta_{2}-\Delta_{3})}{\Gamma(1-\Delta_{2})}\frac{z_{23}^{\Delta_{3}-1}}{\Bar{z}_{32}^{\Delta_{3}-1}}L^{-}[\phi^{-}_{\Delta_{2}+\Delta_{3}-1}]. \label{TotalgluonOPE}
\end{align}
This is precisely what was obtained from an intrinsic analysis in \eqref{IntrinsicTotalgluonOPE}. It should be noted that the above result is for a particular ordering of the operator insertions on the Celestial sphere ($z_{2}>z_{3}>z_{1}$). For the other ordering, $z_{1}>z_{3}>z_{2}$, one would get the same result with an overall negative sign as given in \eqref{IntrinsicTotalgluonOPE2}. 
%\color{red} Also, note that since we are considering the sum of \eqref{gluonOPEMHV} and \eqref{gluonOPEantiMHV} the OPE of $L^{-}[\phi^{-}_{\Delta_{2}}]L^{+}[\phi^{+}_{\Delta_{3}}]$ is the same as the OPE of $L^{+}[\phi^{+}_{\Delta_{2}}]L^{-}[\phi^{-}_{\Delta_{3}}]$ which is evident if one carefully compares \eqref{IntrinsicTotalgluonOPE} and \eqref{TotalgluonOPE}\color{black}.
% This happens because the light-transformed three-point function is defined up to a sign. However, an understanding of this from a kinematic point of view is not known to us and we will address this in future works.
\section{Discussion}
%Let us briefly take a stock of things which we have learnt. 
In the above work, we showed that it is possible to exploit the translation symmetry of the 4D theory to constrain the putative boundary conformal theory without any additional bulk information. We focussed on the useful properties of light transformation of modified Mellin operators. A major tool that helped us understand OPE and the CFT branch of Carrollian CFT correlators is the power law behaviour of correlation functions, which rips the ultra-locality off ordinary Mellin/ modified Mellin amplitudes.

Although we studied light transformation so that the correlation functions have a power law behaviour, a major question that remains open %to which we want to find an answer
is: \textit{What is a good basis of states for Celestial CFT?} Also, given the current correspondence between Celestial CFT and Carrollian CFT, how suitable are these states for studying Carrollian field theories?  
\begin{itemize}

\item We saw in Section \ref{2ptScalar} that light transformation of modified Mellin primaries provides us with a map between the two, thus far, unrelated branches of \textit{3D Carroll CFT}. This map is achieved when considering the near coincident limit of the correlation functions of light-transformed primaries. This map has a natural interpretation in terms of Celestial CFT as can be inferred from \eqref{ModifiedMellin}. Tuning the `null-time' $u$ to zero gives us the usual Mellin transformation of \cite{Pasterski:2017kqt}, and one expects this to hold at the level of correlation functions as well. However, from a Carrollian perspective, this is non-trivial since correlation functions of \textit{2D CFT} branch can never be obtained as a limit of the correlation functions of the \textit{Dirac delta} branch. From a Carrollian viewpoint, one needs to understand what it means to have a correlation function in the near coincident limit while studying the \textit{2D CFT} branch. From another perspective, one can view the CFT branch and the Delta function branch correlators to be different solutions of the same partial differential equations, i.e. the Carroll/BMS Ward identities. This hints that light transformation might be a symmetry of Carrollian CFT. %Whether one can extend all the arguments of this paper to a generic Carrollian conformal field theory is not well understood. We leave these problems for future explorations. Curiously in \cite{Bekaert:2024itn}, it was shown that

\item  We must mention that while finalising the draft of this paper we came across \cite{Ruzziconi:2024zkr} wherein the authors have studied Carrollian amplitudes in a non-trivial dilaton background which leads to breaking of translation invariance and gives non-distributional correlation functions.  However, as we discussed, light transformation naturally gives rise to such correlation functions without breaking translation symmetry. Studying the relationship between light-transformed correlators and field theories with such non-trivial background fields would be interesting. Also, in \cite{Ruzziconi:2024zkr} the authors discuss about differential equations satisfied by Carrollian amplitudes and show that the \textit{2D CFT} branch of Carroll CFT trivially satisfies such differential equations.
%and it was pointed out that the \textit{2D CFT} branch corresponds to magnetic Carroll\cite{Banerjee:2020qjj}.
We wish to understand how light-transformed fields fit into the magnetic Carroll picture. 

\item  While studying free Carrollian fermions in $d=2$ dimensions, it was shown in \cite{Hao:2022xhq, Yu:2022bcp} that there are two types of correlation functions, a power law kind and a Dirac delta kind. A similar conclusion was drawn using generic symmetry arguments in \cite{deBoer:2021jej}. It would be interesting to study such theories within a $3D$ Carrollian framework and understand whether light transformation provides a map between these two kinds of correlation functions.

\item In \cite{Dutta:2022vkg}, it was shown that the OPE of the Carrollian stress tensor with itself consists of contact terms in the leading order. The appearance of such terms could be understood from the fact that Carrollian CFT is an ultra-local theory \cite{Wall:2011hj}. Our analysis in this paper indicates that the OPE of the light-transformed stress tensor will result in OPEs of ordinary 2D CFTs.% A possible way to cure this ultra-locality is to smear the fields in either $z$ or $\Bar{z}$, and it would be interesting to see whether light transformation could be a possible candidate for smearing the fields.

\item It is well known that bulk space-time translations act as global supertranslations on the celestial sphere and owing to the non-local definition of light-transformed operators \eqref{L+}, \eqref{L-} this boils down to the fact that they act as spatial translations on $\mathcal{I}^{\pm}$, as was shown in \ref{PoincareLight}. This implies that under the action of translations, a conformal primary gets mapped to a weight-shifted descendant as we verify in Appendix  \ref{M00action}. Nevertheless, owing to this non-trivial action of translations, one can fix the form of OPE of light-transformed operators modulo a numerical constant. This is novel since, in a usual CFT, one can not fix the leading order singularity of the OPE using the conformal symmetries alone. But having additional symmetries namely bulk translations allows us to do that. 
\item In the context of obtaining OPEs intrinsically, we must mention that the ansatz \eqref{OPEansatz} is motivated by the fact that we are interested in studying the algebra of light-transformed operators only. OPE of light-transformed operators were studied extensively in \cite{De:2022gjn, Hu:2022syq}, and it was shown that in leading order, one obtains not only light-transformed operators but shadow operators as well. It would be interesting to obtain such terms from a purely symmetry-based analysis. 

\section*{Acknowledgement}
We thank Arjun Bagchi, Alok Laddha, Amartya Saha and Daniel Grumiller for their discussions and insights into flat holography. SB would like to thank Chennai Mathematical Institute for its hospitality during his visit there. The grants that support the research of RB are CRG/2020/002035, MTR/ 2022/000795 from SERB, India, DST/IC/Austria/P-9/2021 from DST, India and OeAD Austria; and OPERA grant from BITS Pilani. The authors are grateful to the participants of “Holography, Strings and other fun things” at BITS Pilani (Goa Campus) between 19-23 February 2024, for discussions.

\end{itemize}

\appendix

\section{Ward Identities for Scalars} \label{WardIdentityAppendix}
\subsection{Ward Identity for the two-point function of scalars}\label{WardIdentity2ptAppendix}
\subsubsection*{Ward Identity for translations} 
\begin{enumerate}
    \item Ward Identity for \boldmath{$\delta_{M_{00}}$}
    \unboldmath{}
We will use \eqref{L-M00} and \eqref{L+M00} to get, 
\begin{equation}
\delta_{M_{00}}\langle L^{+}[\phi_{\Delta_{1}}]L^{-}[\phi_{\Delta_{2}}]\rangle=-\frac{i\epsilon_{1}}{1-\Delta_{1}}\partial_{z_{1}}\mathcal{W}(\Delta_{1}+1,\Delta_{2})-\frac{i\epsilon_{2}}{1-\Delta_{2}}\partial_{z_{2}}\mathcal{W}(\Delta_{1},\Delta_{2}+1).
\end{equation}
Using the definition of $\mathcal{W}(\Delta_{1},\Delta_{2})$ \eqref{2ptLight} we get, 
\begin{equation}
\delta_{M_{00}}\langle L^{+}[\phi_{\Delta_{1}}]L^{-}[\phi_{\Delta_{2}}]\rangle= -\frac{\Gamma(\Delta_{1}+\Delta_{2}-1)\Bar{z}^{\Delta_{2}-2}_{12}z^{\Delta_{1}-2}_{21}}{(iu_{12})^{\Delta_{1}+\Delta_{2}-1}}\Big(i\epsilon_{1}+i\epsilon_{2}\Big)=0. \label{M00Ward}
\end{equation} 
%Note that the $\delta_{M_{00}}$ variation hits the functions of $z$ and $\Bar{z}$ in the two-point correlator since $u$ derivatives are re-expressed as $z$ derivatives. Now, the choice of $\epsilon$ only affects the sign in front of $u_{12}$ because $\epsilon$ multiplies $u$ in \eqref{ModifiedMellin}. The two-point function of scalars for arbitrary $\epsilon_{i}$'s would be.This tells us that, even if we were not given the values of the $\epsilon_{i}$'s to begin with, for the light-transformed two-point correlator to satisfy the Ward identity for $M_{00}$ one must have opposite signs for $\epsilon_{1}$ and $\epsilon_{2}$ \footnote{We would like to thank Shamik Banerjee for pointing out this simple yet subtle fact.} \color{red}which is what one would expect from the perspective of energy conservation/four-momentum conservation\color{black}. We will see in further sections that Ward identities for the other global supertranslations will impose this same constraint. 

     \item Ward identity for \boldmath{$\delta_{M_{10}}$} 

We use \eqref{L+M10} and \eqref{L-M10} and get, 
\unboldmath {\begin{align}
    \delta_{M_{10}}\langle L^{+}[\phi_{\Delta_{1}}]L^{-}[\phi_{\Delta_{2}}]\rangle=-i\epsilon_{1}\Big(\frac{z_{1}}{1-\Delta_{1}}\partial_{z_{1}} +  1 \Big)&\mathcal{W}(\Delta_{1}+1,\Delta_{2})\nn \\
    &-\frac{i\epsilon_{2}z_{2}}{1-\Delta_{2}}\partial_{\Bar{z}_{2}}\mathcal{W}(\Delta_{1},\Delta_{2}+1).
\end{align}}
Using \eqref{2ptLight} we get, 
\begin{equation}
   \delta_{M_{10}}\langle L^{+}[\phi_{\Delta_{1}}]L^{-}[\phi_{\Delta_{2}}]\rangle=-\frac{\Gamma(\Delta_{1}+\Delta_{2}-1)\Bar{z}^{\Delta_{2}-2}_{12}z^{\Delta_{1}-2}_{21}}{(iu_{12})^{\Delta_{1}+\Delta_{2}-1}}z_{2}(i\epsilon_{1}+i\epsilon_{2})=0
    \label{M10Ward}
\end{equation}
\item Ward identity for \boldmath{$\delta_{M_{01}}$}
\unboldmath{}

We use \eqref{L+M01} and \eqref{L-M01} to obtain, 
\begin{align}
   \delta_{M_{01}}\langle L^{+}[\phi_{\Delta_{1}}]L^{-}[\phi_{\Delta_{2}}]\rangle=-\frac{i\epsilon_{1}\Bar{z}_{1}}{1-\Delta_{1}}\partial_{z_{1}}\mathcal{W}(\Delta_{1}+1,\Delta_{2})-i\epsilon_{2}\Big(\frac{\Bar{z}_{2}}{1-\Delta_{2}}\partial_{\Bar{z}_{2}}+1 \Big)\mathcal{W}(\Delta_{1},\Delta_{2}+1).
\end{align}
Again using the light-transformed two-point function we get,
    \begin{equation}
       \delta_{M_{01}}\langle L^{+}[\phi_{\Delta_{1}}]L^{-}[\phi_{\Delta_{2}}]\rangle= -\frac{\Gamma(\Delta_{1}+\Delta_{2}-1)\Bar{z}^{\Delta_{2}-2}_{12}z^{\Delta_{1}-2}_{21}}{(iu_{12})^{\Delta_{1}+\Delta_{2}-1}}\Bar{z}_{1}(i\epsilon_{1}+i\epsilon_{2})=0
       \label{M01Ward}
    \end{equation}

\end{enumerate}
Thus we have shown that the light-transformed two-point function has all the translation symmetries. Next, we will verify the Ward identities for Lorentz transformations in respective pairs.
\subsubsection*{Ward Identity for Lorentz transformations}
\begin{itemize}
    \item Ward Identity for $\delta_{L_{-1}}$ and $\delta_{\Bar{L}_{-1}}$

Using \eqref{L+L-1} and \eqref{L-L-1} we get, 
\begin{equation}
    \delta_{L_{-1}}\langle L^{+}[\phi_{\Delta_{1}}]L^{-}[\phi_{\Delta_{2}}]\rangle=\sum_{i=1}^{2}\partial_{z_{i}}\mathcal{W}(\Delta_{1},\Delta_{2}).
\end{equation}
It follows from \eqref{2ptLight} that, 
\begin{equation}
    \delta_{L_{-1}}\langle L^{+}[\phi_{\Delta_{1}}]L^{-}[\phi_{\Delta_{2}}]\rangle=\frac{\Gamma(\Delta_{1}+\Delta_{2}-2)\Bar{z}^{\Delta_{2}-2}_{12}}{(iu_{12})^{\Delta_{1}+\Delta_{2}-2}}(\Delta_{1}-2)(z^{\Delta_{1}-3}_{21}-z^{\Delta_{1}-3}_{21})=0 
     \label{L-1Ward}
\end{equation}
Similarly for, $\delta_{\Bar{L}_{-1}}$ we have, 
\begin{equation}
   \delta_{\Bar{L}_{-1}}\langle L^{+}[\phi_{\Delta_{1}}]L^{-}[\phi_{\Delta_{2}}]\rangle=\sum_{i=1}^{2}\partial_{\Bar{z}_{i}}\mathcal{W}(\Delta_{1},\Delta_{2})
\end{equation}
which evaluates to, 
\begin{equation}
   \delta_{\Bar{L}_{-1}}\langle L^{+}[\phi_{\Delta_{1}}]L^{-}[\phi_{\Delta_{2}}]\rangle=\frac{\Gamma(\Delta_{1}+\Delta_{2}-2)\Bar{z}^{\Delta_{2}-2}_{12}}{(iu_{12})^{\Delta_{1}+\Delta_{2}-2}}(\Delta_{2}-2)(\Bar{z}^{\Delta_{2}-3}_{12}-\Bar{z}^{\Delta_{2}-3}_{12})=0 . \label{Lbar-1Ward}
\end{equation}
These are the only Ward identities that do not depend upon the kinematics of particles which is natural from the perspective of 2D CFT since $\delta_{L_{-1}}$ generate translations along $z$ and our correlators, having a power law behaviour, is expected to satisfy them.  

\item Ward Identity for $\delta_{L_{0}}$ and $\delta_{\Bar{L}_{0}}$

Starting with $\delta_{L_{0}}$ we have from \eqref{L+L0} and \eqref{L-L0},
\begin{align}
    \delta_{L_{0}}\langle L^{+}[\phi_{\Delta_{1}}]L^{-}[\phi_{\Delta_{2}}&]\rangle=\Bigg[\Big(\sum_{i=1}^{2}z_{i}\partial_{z_{i}} +(1-\frac{\Delta_{1}}{2}+\frac{\Delta_{2}}{2})\Big)\mathcal{W}(\Delta_{1},\Delta_{2})\nn \\
    & -\frac{i\epsilon_{1}u_{1}}{2(1-\Delta_{1})}\partial_{z_{1}}\mathcal{W}(\Delta_{1}+1,\Delta_{2})-\frac{i\epsilon_{2}u_{2}}{2(1-\Delta_{2})}\mathcal{W}(\Delta_{1},\Delta_{2}+1)
    \Bigg]
\end{align}
Upon simplification this gives us,
\begin{align}
    \delta_{L_{0}}\langle L^{+}[\phi_{\Delta_{1}}]L^{-}[\phi_{\Delta_{2}}]\rangle=\Bigg[ \Delta_{1}-2+1-&\frac{\Delta_{1}}{2}+\frac{\Delta_{2}}{2}\nn \\
    &-\frac{\Delta_{1}+\Delta_{2}-2}{2u_{12}}(\epsilon_{1}u_{1}+\epsilon_{2}u_{2}) \Bigg]\mathcal{W}(\Delta_{1},\Delta_{2})=0 \label{L0Ward}
\end{align}
since $\epsilon_{1}=1$ and $\epsilon_{2}=-1$. 
Similarly, for $\delta_{\Bar{L}_{0}}$ we have from \eqref{L+Lbar0} and \eqref{L-Lbar0},
\begin{align}
   \delta_{\Bar{L}_{0}}\langle L^{+}[\phi_{\Delta_{1}}]L^{-}&[\phi_{\Delta_{2}}]\rangle=\Bigg[\Big( \sum_{i=1}^{2}\Bar{z}_{i}\partial_{\Bar{z}_{i}}+\frac{\Delta_{1}}{2}+(1-\frac{\Delta_{2}}{2})\Big)\mathcal{W}(\Delta_{1},\Delta_{2})\nn \\
    & -\frac{i\epsilon_{1}u_{1}}{2(1-\Delta_{1})}\partial_{z_{1}}\mathcal{W}(\Delta_{1}+1,\Delta_{2})-\frac{i\epsilon_{2}u_{2}}{2(1-\Delta_{2})}\partial_{\Bar{z}_{2}}\mathcal{W}(\Delta_{1},\Delta_{2}+1).
    \Bigg]
\end{align}
Upon evaluating this using \eqref{2ptLight} we get,
\begin{align}
    \delta_{\Bar{L}_{0}}\langle L^{+}[\phi_{\Delta_{1}}]L^{-}[\phi_{\Delta_{2}}]\rangle= \Bigg[\Delta_{2}-2+1-\frac{\Delta_{2}}{2}+\frac{\Delta_{1}}{2}+\frac{2-\Delta_{1}-\Delta_{2}}{2u_{12}}(\epsilon_{1}u_{1}-\epsilon_{2}u_{2}) \Bigg]\mathcal{W}(\Delta_{1},\Delta_{2})=0.
\end{align}

\end{itemize}
\subsection{Ward Identity for three-point function of scalars}\label{WardIdentity3ptAppendix}
\subsubsection*{Ward Identities for 4D bulk Translations} \label{3ptSupertranslationWard}
 
\begin{enumerate}

 %Again, in the spirit of section \ref{2ptWardIdentity}, one could analyze this result from the perspective of kinematics and see that the $z$ ($\Bar{z}$) derivatives do not act on the function fixed by the $\epsilon_{i}$'s, which is $U$. So \eqref{3ptM00Ward} tells us that for the Ward identity to hold one must have $\epsilon_{1}=\epsilon_{2}=-\epsilon_{3}$ which is what one would have expected for this channel \cite{Pasterski:2017ylz}. 

    \item Ward Identity for \boldmath{$\delta_{M_{00}}$}    
\unboldmath{}

We have from \eqref{L+M00} and \eqref{L-},
    \begin{align}
        \delta_{M_{00}}\langle L^{-}[\phi_{\Delta_1}]L^{-}[\phi_{\Delta_2}]L^{+}[\phi_{\Delta_{3}}] \rangle=-&\frac{i\epsilon_{1}}{1-\Delta_{1}}\partial_{\Bar{z}_{1}}\mathcal{W}(\Delta_{1}+1,\Delta_{2},\Delta_{3})-\frac{i\epsilon_{2}}{1-\Delta_{2}}\partial_{\Bar{z}_{2}}\mathcal{W}(\Delta_{1},\Delta_{2}+1,\Delta_{3})\nn \\
        &-\frac{i\epsilon_{3}}{1-\Delta_{3}}\partial_{z_{3}}\mathcal{W}(\Delta_{1},\Delta_{2},\Delta_{3}+1).
    \end{align}
 Upon using \eqref{3ptScalarLight} this simplifies to,   
\begin{align}
    \delta_{M_{00}}&\langle L^{-}[\phi_{\Delta_1}]L^{-}[\phi_{\Delta_2}]L^{+}[\phi_{\Delta_{3}}] \rangle=4\pi^{2}\Gamma(\beta-3)(-1)^{\Delta_{1}+\Delta_{2}-2}\Bar{z}^{\Delta_{1}-2}_{31}\Bar{z}^{\Delta_{2}-2}_{32}z^{\Delta_{3}-2}_{12}\nn \\
    & \times \int_{z_{1}}^{z_{2}}dw_{3}\frac{(z_{2}-w_{3})^{\Delta_{1}-2}\ (w_{3}-z_{1})^{\Delta_{2}-2}}{(w_{3}-z_{3})^{2-\Delta_{3}} \ U^{\beta-3}} [-z_{1}(i\epsilon_{2}+i\epsilon_{3})+z_{2}(i\epsilon_{1}+i\epsilon_{3})+w_{3}(i\epsilon_{2}-i\epsilon_{1})]=0 \label{3ptM00Ward}
\end{align}
 where we have used $\epsilon_{1}=\epsilon_{2}=-1$ and $\epsilon_{3}=1$ from \eqref{3ptScalarMellin}.

\item Ward Identity for \boldmath{$\delta_{M_{10}}$}

 \unboldmath{}
We have from \eqref{L+M10} and \eqref{L-M10}
\begin{align}
    \delta_{M_{10}}\langle L^{-}[\phi_{\Delta_1}]L^{-}[\phi_{\Delta_2}]L^{+}[\phi_{\Delta_{3}}] \rangle=&-\frac{i\epsilon_{1}z_{1}}{1-\Delta_{1}}\partial_{\Bar{z}_{1}}\mathcal{W}(\Delta_{1}+1,\Delta_{2},\Delta_{3})-\frac{i\epsilon_{2}z_{2}}{1-\Delta_{2}}\partial_{\Bar{z}_{2}}\mathcal{W}(\Delta_{1},\Delta_{2}+1,\Delta_{3})\nn \\
    & -i\epsilon_{3}\Big( \frac{z_{3}}{1-\Delta_{3}}\partial_{z_{3}}+1\Big)\mathcal{W}(\Delta_{1},\Delta_{2},\Delta_{3}+1). 
\end{align}
Using the light-transformed three-point function and the proper signs for the $\epsilon_{i}$'s we get,
\begin{align}
    \delta_{M_{10}}&\langle L^{-}[\phi_{\Delta_1}]L^{-}[\phi_{\Delta_2}]L^{+}[\phi_{\Delta_{3}}] \rangle=4\pi^{2}\Gamma(\beta-3)(-1)^{\Delta_{1}+\Delta_{2}-2}\Bar{z}^{\Delta_{1}-2}_{31}\Bar{z}^{\Delta_{2}-2}_{32}z^{\Delta_{3}-2}_{12}\nn \\
    & \times \int_{z_{1}}^{z_{2}}dw_{3}\frac{(z_{2}-w_{3})^{\Delta_{1}-2}\ (w_{3}-z_{1})^{\Delta_{2}-2}}{(w_{3}-z_{3})^{2-\Delta_{3}} \ U^{\beta-3}}i[z_{1}z_{2}(\epsilon_{1}-\epsilon_{2})-z_{1}w_{3}(\epsilon_{1}+\epsilon_{3})+z_{2}w_{3}(\epsilon_{2}+\epsilon_{3})]=0 .\label{3ptM10Ward}
\end{align}
 \item Ward Identity for \boldmath{$\delta_{M_{01}}$} \unboldmath{}

We have from equations \eqref{L+M10} and \eqref{L-M10} 
\begin{align}
    \delta_{M_{01}}\langle L^{-}[\phi_{\Delta_1}]L^{-}[\phi_{\Delta_2}]&L^{+}[\phi_{\Delta_{3}}] \rangle=-i\epsilon_{1}\Big(\frac{\Bar{z}_{1}}{1-\Delta_{1}}\partial_{\Bar{z}_{1}} + 1 \Big)\mathcal{W}(\Delta_{1}+1,\Delta_{2},\Delta_{3})\nn \\
    & -i\epsilon_{2}\Big( \frac{\Bar{z}_{2}}{1-\Delta_{2}}\partial_{\Bar{z}_{2}} + 1  \Big)\mathcal{W}(\Delta_{1},\Delta_{2}+1,\Delta_{3})-\frac{i\epsilon_{3}\Bar{z}_{3}}{1-\Delta_{3}}\mathcal{W}(\Delta_{1},\Delta_{2},\Delta_{3}+1).
\end{align}
Again using \eqref{3ptScalarLight} and the $\epsilon_{i}$'s we have,
\begin{align}
   \delta_{M_{01}}&\langle L^{-}[\phi_{\Delta_1}]L^{-}[\phi_{\Delta_2}]L^{+}[\phi_{\Delta_{3}}] \rangle=4\pi^{2}\Gamma(\beta-3)(-1)^{\Delta_{1}+\Delta_{2}-2}\Bar{z}^{\Delta_{1}-2}_{31}\Bar{z}^{\Delta_{2}-2}_{32}z^{\Delta_{3}-2}_{12}\nn \\
    & \times \int_{z_{1}}^{z_{2}}dw_{3}\frac{(z_{2}-w_{3})^{\Delta_{1}-2}\ (w_{3}-z_{1})^{\Delta_{2}-2}}{(w_{3}-z_{3})^{2-\Delta_{3}} \ U^{\beta-3}}i[\Bar{z}_{3}z_{2}(\epsilon_{1}+\epsilon_{2})+\Bar{z}_{3}w_{3}(\epsilon_{2}-\epsilon_{1})-\Bar{z}_{3}z_{1}(\epsilon_{2}+\epsilon_{3})]=0. \label{3ptM01Ward}
\end{align}

\end{enumerate}
This shows that the light-transformed three-point function satisfies the Ward identities for 4D Bulk translations. In the next section, we will show that they also satisfy the Ward identity for Lorentz transformations. 

\subsubsection*{Ward Identity for Lorentz transformations}\label{3ptL-1}
\begin{enumerate}

\item Ward Identity for \boldmath{$\delta_{L_{0}}$} and \boldmath{$\delta_{\Bar{L}_{0}}$} \unboldmath{}
We use \eqref{L+L0} and \eqref{L-L0} we have in terms of $\mathcal{W}(\Delta_{1},\Delta_{2},\Delta_{3})$, 
\begin{align}
    \delta_{L_{0}}\langle L^{-}[\phi_{\Delta_1}]&L^{-}[\phi_{\Delta_2}]L^{+}[\phi_{\Delta_{3}}] \rangle=\Bigg[\bigg(\sum_{i=1}^{3}z_{i}\partial_{z_{i}}+\frac{\Delta_{1}}{2}+\frac{\Delta_{2}}{2}+(1-\frac{\Delta_{3}}{2})\bigg)\mathcal{W}(\Delta_{1},\Delta_{2},\Delta_{3})\nn \\
    & -\frac{i\epsilon_{1}u_{1}}{2(1-\Delta_{1})}\partial_{\Bar{z}_{1}}\mathcal{W}(\Delta_{1}+1,\Delta_{2},\Delta_{3})-\frac{i\epsilon_{2}u_{2}}{2(1-\Delta_{2})}\partial_{\Bar{z}_{2}}\mathcal{W}(\Delta_{1},\Delta_{2}+1,\Delta_{3})\nn \\
    &-\frac{i\epsilon_{3}u_{3}}{2(1-\Delta_{3})}\partial_{z_{3}}\mathcal{W}(\Delta_{1},\Delta_{2},\Delta_{3}+1)\Bigg].
\end{align}
We re-write the $\partial_{z_{3}}\mathcal{W}(\Delta_{1},\Delta_{2},\Delta_{3})$ and get the following form, 
\begin{align}
    z_{3}\partial_{z_{3}}\mathcal{W}(\Delta_{1},\Delta_{2},\Delta_{3}&)=-\bigg(w_{3}\frac{\mathcal{M}(\Delta_{1},\Delta_{2},\Delta_{3})}{(w_{3}-z_{3})^{2-\Delta_{3}}}\bigg)\Bigg|_{z_{1}}^{z_{2}} \nn \\
    &-(1-\Delta_{3})\mathcal{W}(\Delta_{1},\Delta_{2},\Delta_{3})+\int_{z_{1}}^{z_{2}}dw_{3}\frac{w_{3}\partial_{w_{3}}\mathcal{M}(\Delta_{1},\Delta_{2},\Delta_{3})}{(w_{3}-z_{3})^{2-\Delta_{3}}}. \label{DivergenceCancellationL0}
\end{align}
This expression can be proved by performing integration by parts. Using the above-mentioned result we get the following simplification, 
\begin{align}
    \delta_{L_{0}}\langle L^{-}[\phi_{\Delta_1}]L^{-}[\phi_{\Delta_2}]L^{+}[\phi_{\Delta_{3}}] \rangle=\bigg[\beta-6+4-\beta+2+\frac{\beta}{2}-\frac{\beta}{2} \bigg]\mathcal{W}(\Delta_{1},\Delta_{2},\Delta_{2})=0. \label{3ptL0Ward}
\end{align}
Next we calculate the Ward identity for $\delta_{\Bar{L}_{0}}$ using \eqref{L+Lbar0} and \eqref{L-Lbar0},
\begin{align}
    \delta_{\Bar{L}_{0}}\langle L^{-}[\phi_{\Delta_1}]&L^{-}[\phi_{\Delta_2}]L^{+}[\phi_{\Delta_{3}}] \rangle= \Bigg[\bigg(\sum_{i=1}^{3} \Bar{z}_{i}\partial_{\Bar{z}_{i}}+2+\frac{\Delta_{3}-\Delta_{1}-\Delta_{2}}{2}\bigg)\mathcal{W}(\Delta_{1},\Delta_{2},\Delta_{3}) \nn \\
    & -\frac{i\epsilon_{1}u_{1}}{2(1-\Delta_{1})}\partial_{\Bar{z}_{1}}\mathcal{W}(\Delta_{1}+1,\Delta_{2},\Delta_{3})-\frac{i\epsilon_{2}u_{2}}{2(1-\Delta_{2})}\partial_{\Bar{z}_{2}}\mathcal{W}(\Delta_{1},\Delta_{2}+1,\Delta_{3})\nn \\
    &-\frac{i\epsilon_{3}u_{3}}{2(1-\Delta_{3})}\partial_{z_{3}}\mathcal{W}(\Delta_{1},\Delta_{2},\Delta_{3}+1)\Bigg].
\end{align}
Simplifying this we get, 
\begin{equation}
 \delta_{\Bar{L}_{0}}\langle L^{-}[\phi_{\Delta_1}]L^{-}[\phi_{\Delta_2}]L^{+}[\phi_{\Delta_{3}}] \rangle=\bigg[\Delta_{1}+\Delta_{2}-4+2-\frac{\Delta_{1}+\Delta_{2}-\Delta_{3}}{2}+2-\frac{\beta}{2} \bigg]\mathcal{W}(\Delta_{1},\Delta_{2},\Delta_{3})=0 \label{3ptLbar0Ward}
\end{equation}

\item Ward Identity for \boldmath{$\delta_{L_{1}}$} and \boldmath{$\delta_{\Bar{L}_{1}}$} \unboldmath{}

Using \eqref{L+L1} and \eqref{L-L1} we get, 
\begin{align}
    \delta_{\Bar{L}_{1}}\langle L^{-}[\phi_{\Delta_1}]L^{-}[\phi_{\Delta_2}]&L^{+}[\phi_{\Delta_{3}}] \rangle=\bigg(\sum_{i=1}^{3} z_{i}^{2}\partial_{z_{i}}+\Delta_{1}z_{1}+\Delta_{2}z_{2} +(2-\Delta_{3})z_{3}\bigg)\mathcal{W}(\Delta_{1},\Delta_{2},\Delta_{3})\nn \\
    &-\frac{i\epsilon_{1}u_{1}}{1-\Delta_{1}}z_{1}\partial_{\Bar{z}_{1}}\mathcal{W}(\Delta_{1}+1,\Delta_{2},\Delta_{3})-\frac{i\epsilon_{2}u_{2}}{1-\Delta_{2}}z_{2}\partial_{\Bar{z}_{2}}\mathcal{W}(\Delta_{1},\Delta_{2}+1,\Delta_{3})\nn \\
    &-i\epsilon_{3}u_{3}\bigg(\frac{z_{3}}{1-\Delta_{3}}\partial_{z_{3}}+1 \bigg)\mathcal{W}(\Delta_{1},\Delta_{2},\Delta_{3}+1)
\end{align}
 We will re-write the $z_{3}^{2}\partial_{z_{3}}+(2-\Delta_{3})$ term as , 
\begin{align}
    (z_{3}^{2}\partial_{z_{3}}+2-\Delta_{3})\mathcal{W}(\Delta_{1},\Delta_{2},\Delta_{3})=-&\bigg(w_{3}^{2}\frac{\mathcal{M}(\Delta_{1},\Delta_{2},\Delta_{3})}{(w_{3}-z_{3})^{2-\Delta_{3}}}\bigg)\Bigg|_{z_{1}}^{z_{2}}\nn\\
    &+\int_{z_{1}}^{z_{2}}dw_{3}\frac{(w_{3}^{2}\partial_{w_{3}}+\Delta_{3}w_{3})\mathcal{M}(\Delta_{1},\Delta_{2},\Delta_{3})}{(w_{3}-z_{3})^{2-\Delta_{3}}}. \label{DivergenceCancellationL1}
\end{align}
Substituting this in the equation for Ward identity and using the three-point function \eqref{3ptScalarLight} one gets, 
\begin{align}
    \delta_{\Bar{L}_{1}}\langle L^{-}[\phi_{\Delta_1}]L^{-}[\phi_{\Delta_2}]L^{+}[\phi_{\Delta_{3}}] \rangle=\Bigg[z_{1}\big(\beta-4+4-\beta   \big)+&z_{2}\big(\beta-4+4-\beta \big)\nn \\
    &+z_{3}\big(\beta-4 +4-\beta \big)  \Bigg]=0. \label{3ptL1Ward}
\end{align}
Moving on we check the Ward identity for $\delta_{\Bar{L}_{1}}$ using \eqref{L+Lbar1} and \eqref{L-Lbar1} and get, 
\begin{align}
    &\delta_{\Bar{L}_{1}}\langle L^{-}[\phi_{\Delta_1}]L^{-}[\phi_{\Delta_2}]L^{+}[\phi_{\Delta_{3}}] \rangle=\Bigg[\bigg(\sum_{i=1}^{3}\Bar{z}_{i}^{2}\partial_{\Bar{z}_{i}}+4-\Delta_{1}-\Delta_{2}+\Delta_{3}\bigg)\mathcal{W}(\Delta_{1},\Delta_{2},\Delta_{3})\nn \\
    &-i\epsilon_{1}u_{1}\bigg(\frac{\Bar{z}_{1}}{1-\Delta_{1}}\partial_{z_{1}}+1\bigg)\mathcal{W}(\Delta_{1}+1,\Delta_{2},\Delta_{3})-i\epsilon_{2}u_{2}\bigg(\frac{\Bar{z}_{2}}{1-\Delta_{2}}\partial_{z_{2}}+1\bigg)\mathcal{W}(\Delta_{1},\Delta_{2}+1,\Delta_{3})\nn \\
    &-\frac{i\epsilon_{3}u_{3}}{1-\Delta_{3}}\Bar{z}_{3}\partial_{z_{3}}\mathcal{W}(\Delta_{1},\Delta_{2},\Delta_{3}+1) \Bigg].
\end{align}
We simplify this using \eqref{3ptScalarLight} and get, 
\begin{align}
    \delta_{\Bar{L}_{1}}\langle L^{-}[\phi_{\Delta_1}]L^{-}[\phi_{\Delta_2}]L^{+}[\phi_{\Delta_{3}}] \rangle=\Bigg[\Bar{z}_{1}\bigg(\Delta_{1}-2+2-\Delta_{2}\bigg) + & \Bar{z}_{2}\bigg(\Delta_{2}-2+2-\Delta_{2}\bigg)\nn \\
    &+\Bar{z}_{1}\bigg(\beta-4+4-\beta \bigg) \Bigg]=0. \label{3ptLbar1Ward}
\end{align}
\end{enumerate}
Thus the light-transformed three-point function of scalars satisfies all the Ward identities of the Poincar\'e transformations.

\section{Non-triviality of the action of \texorpdfstring{$\delta_{M_{00}}$}{\265}} \label{M00action}
Fields that appear as $z$ (respectively $\Bar{z}$) derivatives of primary fields are referred to as \textit{descendants} \cite{Blumenhagen2009} whose weights are shifted by a factor of $\frac{1}{2}$. In a 3D Carrollian CFT one would expect that action of $\delta_{M_{00}}$ give us \textit{$\partial_{u}$-descendent} \cite{Mason:2023mti}\footnote{In a generic 2D CFT one has a highest weight state corresponding to which one obtains descendants by the application of $L_{-1}$ or $\partial_{z}$. It is not clear what is the highest weight representation of the translations.} Looking at \eqref{L+M00} and \eqref{L-M00} one would be compelled to think that the action of $\delta_{M_{00}}$ has resulted in a \textit{$\partial_{u}$ descendant} of a $z$ descendant. We show that this is exactly the case by studying how $\delta_{M_{00}}L^{+}[\phi_{a,\mathfrak{h}}]$ transforms under $\delta_{L_{0}}$. We start with \eqref{L+M00} and \eqref{L+L0},
\begin{align}
    \delta_{L_{0}}\delta_{M_{00}}L^{+}[\phi_{a,\mathfrak{h}}]=-\partial_{u}\Bigg[\bigg(z\partial_{z} +\frac{1}{2}-h\bigg)L^{+}[\phi_{a,\mathfrak{h}}] +u\partial_{u}L^{+}[\phi_{a,\mathfrak{h}+\frac{1}{2}}]\bigg],
\end{align}
straightforward calculation gives us, 
\begin{align}
    \delta_{L_{0}}\delta_{M_{00}}L^{+}[\phi_{a,\mathfrak{h}}]=-i\Bigg[\bigg(z\partial_{z} +\frac{3}{2}-h \bigg)\delta_{M_{00}}L^{+}[\phi_{a,\mathfrak{h}}]+u\partial_{u}\delta_{M_{00}}L^{+}[\phi_{a,\mathfrak{h}+\frac{1}{2}}]\Bigg]. \label{ScalingofL+M00}
\end{align}
The first term tells us that the conformal weight of this operator is $(\frac{3}{2}-h,\Bar{h})$ which shows, along with the fact that light-transformed operators have conformal weights $(1-h,\bar{h})$, that $\delta_{M_{00}}$ shifts the weight of operators by $\frac{1}{2}$. 

To confirm whether this is a descendant or not we study the variation of $\delta_{M_{00}}L^{+}[\phi_{a,\mathfrak{h}}]$ under $\delta_{L_{1}}$, and get,
\begin{align}
    \delta_{L_{1}}\delta_{M_{00}}L^{+}&[\phi_{a,\mathfrak{h}}]=\bigg(z^{2}\partial_{z}+(3-2h)z+(1-2h)\bigg)(\delta_{M_{00}}L^{+}[\phi_{a,\mathfrak{h}}])\nn \\
    &-i\epsilon u \bigg( \frac{1}{-2h}\Big( \frac{-i \epsilon}{1-2h}\partial_{z}L^{+}[\phi_{a,\mathfrak{h}+1}]-\frac{i\epsilon}{1-2h}z\partial_{z}L^{+}[\phi_{a,\mathfrak{h}+1}]\Big) 
    \frac{-\epsilon}{1-2h}\partial_{z}L^{+}[\phi_{a,\mathfrak{h}+1}] \bigg).
\end{align}
Now, one can verify that at $(u=0,z=0,\bar{z}=0)$ the value of $ \delta_{L_{1}}\delta_{M_{00}}L^{+}[\phi_{a,\mathfrak{h}}]\neq 0$. This shows that the action of $\delta_{M_{00}}$ gives us descendants instead of primaries. 
%We are interested in analysing the term in \color{red} red \color{black} since that tells us about the conformal weights of the primary. We know from \eqref{L+L0} that $L^{+}[\phi_{a,\mathfrak{h}}]$ has conformal weights $(1-h,\Bar{h})$, thus the primary $L^{+}[\phi_{a,\mathfrak{h}+\frac{1}{2}}]$ should have conformal weights $(\frac{1}{2}-h,\Bar{h})$. Now, if the action of $\delta_{M_{00}}$ only raised the weight of the modified Mellin primary, then the factor in red should have been $\frac{1}{2}-h$ which is not the case. This tells us that $\delta_{M_{00}}$ increases the conformal weights of the Light transformed primary by $\frac{1}{2}$ which happens because of the presence of $z$ descendant (basically we can write $\frac{3}{2}-h=1-h+\frac{1}{2}$).

\section{Soft Factorization of light-transformed Four Point correlator} \label{SoftLightFour}
 We begin with an anti-MHV Mellin-transformed four-point correlator. 
\begin{align}
    &\langle\phi^{+}_{\Delta_{1}}\phi^{+}_{\Delta_{2}}\phi^{-}_{\Delta_{3}}\phi^{-}_{\Delta_{4}}\rangle=\frac{1}{4}\frac{\Gamma(\sum_{i=1}^{4}\Delta_{i}-4)}{\Big[i \epsilon_{4}\big( \frac{\Bar{z}_{24}z_{43}}{\Bar{z}_{12}z_{13}}u_{1}+\frac{\Bar{z}_{14}z_{43}}{\Bar{z}_{12}z_{32}}u_{2}+\frac{\Bar{z}_{24}z_{14}}{\Bar{z}_{32}z_{13}}u_{3}+u_{4}\big) \Big]^{\sum_{i=1}^{4}\Delta_{i}-4}} \frac{\Bar{z}_{12}^{3}}{\Bar{z}_{23}\Bar{z}_{34}\Bar{z}_{41}} \nn \\
    & \times \bigg(\frac{\epsilon_{4}\Bar{z}_{24}z_{43}}{\epsilon_{1}\Bar{z}_{12}z_{13}}\bigg)^{\Delta_{1}} \bigg(\frac{\epsilon_{4}\Bar{z}_{14}z_{43}}{\epsilon_{2}\Bar{z}_{12}z_{32}}\bigg)^{\Delta_{2}} \bigg(\frac{\epsilon_{4}\Bar{z}_{24}z_{14}}{\epsilon_{3}\Bar{z}_{32}z_{13}}\bigg)^{\Delta_{3}-2} \Theta \bigg(\frac{\epsilon_{4}\Bar{z}_{24}z_{43}}{\epsilon_{1}\Bar{z}_{12}z_{13}}\bigg) \Theta \bigg(\frac{\epsilon_{4}\Bar{z}_{14}z_{43}}{\epsilon_{2}\Bar{z}_{12}z_{32}}\bigg) \Theta \bigg(\frac{\epsilon_{4}\Bar{z}_{24}z_{14}}{\epsilon_{3}\Bar{z}_{32}z_{13}}\bigg) \nn \\
    & \times \delta(\Bar{z}_{12}\Bar{z}_{34}z_{14}z_{23}-\Bar{z}_{14}\Bar{z}_{23}z_{12}z_{34}) \label{4ptAnti-MHVMellin}
\end{align}
Light transforming this amplitude gives us, 
\begin{align}
   & \langle L^{+}[\phi^{+}_{\Delta_{1}}]L^{+}[\phi^{+}_{\Delta_{2}}]L^{-}[\phi^{-}_{\Delta_{3}}]L^{-}[\phi^{-}_{\Delta_{4}}]\rangle= \frac{1}{4}\int \frac{dw_{1} dw_{2} d\Bar{w}_{3} d\Bar{w}_{4}}{(w_{1}-z_{1})^{1-\Delta_{1}}(w_{2}-z_{2})^{\Delta_{2}-1}(\Bar{w}_{3}-\Bar{z}_{3})^{\Delta_{3}-1}(\bar{w}_{4}-z_{4})^{\Delta_{4}-1}}\nn \\ &\frac{\Gamma(\sum_{i=1}^{4}\Delta_{i}-4)}{\Big[i \epsilon_{4}\big( \frac{(\Bar{z}_{2}-\Bar{w}_{4})z_{43}}{\Bar{z}_{12}(z_{1}-w_{3})}u_{1}+\frac{(\Bar{z}_{1}-\Bar{w}_{4})z_{43}}{\Bar{z}_{12}(z_{3}-w_{2})}u_{2}+\frac{(\Bar{z}_{2}-\Bar{w}_{4})(w_{1}-z_{4})}{(\Bar{w}_{3}-z_{2})(w_{1}-z_{3})}u_{3} +u_{4} \big) \Big]^{\sum_{i=1}^{4}\Delta_{i}-4}} \frac{\Bar{z}_{12}^{3}}{(\Bar{z}_{2}-\Bar{w}_{3})(\Bar{w}_{3}-\Bar{w}_{4})(\Bar{w}_{4}-\Bar{z}_{1})}\nn \\
   &  \times \bigg( \frac{\epsilon_{4}(\Bar{z}_{2}-\Bar{w}_{4})z_{43}}{\epsilon_{1}\Bar{z}_{12}(z_{1}-w_{3})} \bigg)^{\Delta_{1}} \bigg( \frac{\epsilon_{4}(\Bar{z}_{1}-\Bar{w}_{4})z_{43}}{\epsilon_{2}\Bar{z}_{12}(z_{3}-w_{2})} \bigg)^{\Delta_{2}} \bigg( \frac{\epsilon_{4}(\Bar{z}_{2}-\Bar{w}_{4})(w_{1}-z_{4})}{\epsilon_{3}(\Bar{w}_{3}-z_{2})(w_{1}-z_{3})}
 \bigg)^{\Delta_{3}-2} \nn \\
 & \times  \Theta \bigg( \frac{\epsilon_{4}(\Bar{z}_{2}-\Bar{w}_{4})z_{43}}{\epsilon_{1}\Bar{z}_{12}(z_{1}-w_{3})} \bigg) \Theta \bigg( \frac{\epsilon_{4}(\Bar{z}_{1}-\Bar{w}_{4})z_{43}}{\epsilon_{2}\Bar{z}_{12}(z_{3}-w_{2})} \bigg) \Theta \bigg( \frac{\epsilon_{4}(\Bar{z}_{2}-\Bar{w}_{4})(w_{1}-z_{4})}{\epsilon_{3}(\Bar{w}_{3}-z_{2})(w_{1}-z_{3})}
 \bigg) \nn \\
 &\times \delta(\Bar{z}_{12}(\Bar{w}_{3}-\Bar{w}_{4})(w_{1}-z_{4})(w_{2}-z_{3})-(\Bar{z}_{1}-\Bar{w}_{4})(\Bar{z}_{2}-\Bar{w}_{3})(w_{1}-w_{2})z_{34}). \label{4ptAnti-MHVLight}
\end{align}
Taking the soft limit as in section \ref{SoftGluon} and following the analysis of \cite{Pate:2019mfs, Banerjee:2019prz}, one obtains, 
\begin{align}
    &\lim_{\Delta_{3}-1 \rightarrow 0} (\Delta_{3}-1)\langle L^{+}[\phi^{+}_{\Delta_{1}}]L^{+}[\phi^{+}_{\Delta_{2}}]L^{-}[\phi^{-}_{\Delta_{3}}]L^{-}[\phi^{-}_{\Delta_{4}}]\rangle=\frac{1}{4} \Gamma(\Delta_{1}+\Delta_{2}+\Delta_{4}-3)z_{41}^{\Delta_{1}-1}z_{42}^{\Delta_{2}-1}\nn \\
    & \int d\Bar{w}_{3} d\Bar{w}_{4} \frac{(\Bar{w}_{4}-\Bar{w}_{2})}{(\Bar{w}_{3}-\Bar{w}_{4})(\Bar{z}_{2}-\Bar{w}_{3})} 
    \Big[ \frac{(\Bar{w}_{4}-\Bar{z}_{2})^{\Delta_{1}-2}(\Bar{z}_{1}-\Bar{w}_{4})^{\Delta_{2}-1}(\Bar{w}_{4}-z_{4})^{\Delta_{4}-1}}{[i\big((\Bar{z}_{2}-\Bar{w}_{4})u_{1}+(\Bar{w}_{4}-\Bar{z}_{2})u_{2}+\bar{z}_{12}u_{4} \big)]^{\Delta_{1}+\Delta_{2}+\Delta_{4}-3}}\Big]
\end{align}
here one can note that the argument inside the integral is the soft factor, one would expect to get from the conformally soft theorem, multiplied by the three-point function. So, what we have obtained is, in some sense, a light transformation of the conformally soft theorem. Whether one can obtain a soft factorization of these operators or not is something that we leave for future perusal.

\section{Algebra of Light transformed generators} \label{SectionAlgebra}
%Before we check the Ward identities for light-transformed correlators, we will ensure that the global BMS algebra is realized faithfully on these operators.
We derive the algebra satisfied by the generators derived in equations \eqref{L+M00} through \eqref{L-Lbar1}. Here we give a few of the commutation relations satisfied by the generators, one can check that the algebra of generators are identical to the $\mathcal{BMS}_{4}$ algebra.  
%Since two types of light transformations are defined in \eqref{L+} and \eqref{L-}, we will have two sets of realizations of the algebra. For brevity, we denote the light-transformed generators with an extra tilde, $\Tilde{\delta}$. 
After some straightforward calculation, one finds, 
\begin{align}
    & [\delta_{L_{-1}},\delta_{L_{0}}]L^{+}[\phi_{a,\mathfrak{h}}] =-\delta_{L_{-1}}L^{+}[\phi_{a,\mathfrak{h}}] \qquad [\delta_{L_{-1}},\delta_{L_{0}}]L^{-}[\phi_{a,\mathfrak{h}}] =-\delta_{L_{-1}}L^{-}[\phi_{a,\mathfrak{h}}]\\
    & [\delta_{L_{-1}},\delta_{L_{1}}]L^{+}[\phi_{a,\mathfrak{h}}] =-2\delta_{L_{0}}L^{+}[\phi_{a,\mathfrak{h}}] \qquad [\delta_{L_{-1}},\delta_{L_{1}}]L^{-}[\phi_{a,\mathfrak{h}}] =-2\delta_{L_{0}}L^{-}[\phi_{a,\mathfrak{h}}] \\
    &[\delta_{L_{0}},\delta_{L_{1}}]L^{+}[\phi_{a,\mathfrak{h}}] =-\delta_{L_{1}}L^{+}[\phi_{a,\mathfrak{h}}] \qquad [\delta_{L_{0}},\delta_{L_{1}}]L^{-}[\phi_{a,\mathfrak{h}}] =-\delta_{L_{1}}L^{-}[\phi_{a,\mathfrak{h}}]\\
    &[\delta_{L_{-1}},\delta_{M_{00}}]L^{+}[\phi_{a,\mathfrak{h}}] =0\qquad [\delta_{L_{-1}},\delta_{M_{00}}]L^{-}[\phi_{a,\mathfrak{h}}] =0 \\
    & [\delta_{L_{-1}},\delta_{M_{10}}]L^{+}[\phi_{a,\mathfrak{h}}] =-\delta_{M_{00}}L^{+}[\phi_{a,\mathfrak{h}}]\qquad [\delta_{L_{-1}},\delta_{M_{00}}]L^{-}[\phi_{a,\mathfrak{h}}] =-\delta_{M_{00}}L^{+}[\phi_{a,\mathfrak{h}}] \\
    & [\delta_{L_{-1}},\delta_{M_{01}}]L^{+}[\phi_{a,\mathfrak{h}}] =0\qquad [\delta_{L_{-1}},\delta_{M_{01}}]L^{-}[\phi_{a,\mathfrak{h}}] =0  \\
    & [\delta_{L_{-1}},\delta_{M_{11}}]L^{+}[\phi_{a,\mathfrak{h}}] =-\delta_{M_{01}}L^{+}[\phi_{a,\mathfrak{h}}]\qquad [\delta_{L_{-1}},\delta_{M_{11}}]L^{-}[\phi_{a,\mathfrak{h}}] =-\delta_{M_{01}}L^{+}[\phi_{a,\mathfrak{h}}]  \\
    & [\delta_{L_{0}},\delta_{M_{00}}]L^{+}[\phi_{a,\mathfrak{h}}] =\frac{1}{2}\delta_{M_{00}}L^{+}[\phi_{a,\mathfrak{h}}] 
\end{align}

\bibliographystyle{JHEP}
\bibliography{main_with_corrections_.bib}

\end{document}